\documentclass[USletter,11pt]{emulateapj}
\usepackage{epsfig}
\usepackage{natbib}
\usepackage{graphics}
\usepackage{graphicx}
\usepackage{multirow}
\usepackage{amssymb}
\usepackage{color}
\usepackage{boxedminipage}
\usepackage[normalem]{ulem}

\def \kms {{\rm km s$^{-1}$}}
\def \kmsMpc {{\rm km s$^{-1}$ Mpc$^{-1}$}}

\shorttitle{Cosmography of the Local Universe}
\shortauthors{Courtois et al.}

\begin{document}

\title{Cosmography of the Local Universe}

\author{H\'el\`ene M. Courtois$^{1,2}$}
\affil{$^1$University of Lyon; UCB Lyon 1/CNRS/IN2P3; IPN Lyon, France} 
\affil{$^2$Institute for Astronomy (IFA), University of Hawaii, 2680 Woodlawn Drive, HI 96822, USA}
\email{h.courtois@ipnl.in2p3.fr}
\author{Daniel Pomar\`ede$^3$}
\affil{$^3$CEA/IRFU, Saclay, 91191 Gif-sur-Yvette, France} 
\author{R. Brent Tully$^2$}
\affil{$^1$Institute for Astronomy (IFA), University of Hawaii, 2680 Woodlawn Drive, HI 96822, USA }
\author{Yehuda Hoffman$^4$}
\affil{$^4$Racah Institute of Physics, Hebrew University, Jerusalem 91904, Israel}
\and
\author{Denis Courtois${^5}$}
\affil{Lyc\'ee international, 38300 Nivolas-vermelle, France}

\begin{abstract}
The large scale structure of the universe is a complex web of clusters, filaments, and voids.  Its properties are informed by galaxy redshift surveys and measurements of peculiar velocities.  Wiener Filter reconstructions recover three-dimensional velocity and total density fields.  The richness of the elements of our neighborhood are revealed with sophisticated visualization tools.  A key component of this paper is an accompanying movie.  The ability to translate and zoom helps the viewer follow structures in three dimensions and grasp the relationships between features on different scales while retaining a sense of orientation.  The ability to dissolve between scenes provides a technique for comparing different information, for example, the observed distribution of galaxies, smoothed representations of the distribution accounting for selection effects, observed peculiar velocities, smoothed and modeled representations of those velocities, and inferred underlying density fields. The agreement between the large scale structure seen in redshift surveys and that inferred from reconstructions based on the radial peculiar velocities of galaxies strongly supports the standard model of cosmology where structure forms from gravitational instabilities and galaxies form at the bottom of potential wells.
\end{abstract}


\section{Introduction}
\label{sec:intro}

Our goal with this paper is to nurture an enhanced familiarity with the nearby large scale structure in the distribution of galaxies.  We explore the use of video.  Motion helps the eye unravel complex patterns.  We consider the video to be our primary product.  The figures in this article are scenes from the video while the text is mostly a story line connecting and explaining the figures.  The video is available at http://irfu.cea.fr/cosmography.

Information on the distribution of galaxies in redshift space is by now quite extensive and information on the peculiar motions of galaxies is improving rapidly.  Detailed maps of motions are inputs to the translation from redshift space to physical space and constrain the underlying distribution of (mostly dark) matter.  The simple question can be asked if the distribution of observed galaxies and the distribution of matter inferred from galaxy motions agree.  They should, if galaxies are reasonable tracers of the ensemble of matter, and if our procedures are valid for the interpretation of motions in terms of the matter distribution.  The video provides visual comparisons which, although qualitative, are strongly supportive of these propositions.

This paper touches lightly on the following issues: the observed distribution of galaxies in redshift space, the observed radial peculiar velocities of the small fraction of galaxies with accurately known distances, the inferred three-dimensional motions of all the galaxies from a Wiener Filter reconstruction, and the inferred density distribution from these motions.  The details regarding all these steps are discussed in other publications.  Our purpose here is to highlight the usefulness -- the importance -- of visual representations as an arm of this research enterprise.  We call what we are doing `cosmography'.  We argue that maps, with names for features, promote a familiarity and specificity that contributes to physical understanding.

The most fundamental interests in our overall Cosmic Flows program are in the current distribution of matter and in the initial conditions that gave rise to what has evolved.  Our markers are galaxies.  We measure crude three-dimensional (3D) locations from their systematic velocities and sky positions (distorted from true positions by peculiar velocities) and we can estimate their baryonic masses from their luminosities.  Unfortunately this information degrades with distance and can be lacking in regions of heavy obscuration.  In the case of redshift surveys, low luminosity galaxies are mostly lost, with progressively brighter galaxies lost at larger distances, until the survey reaches an effective limit due to the exponential cutoff of the galaxy luminosity function.  
Distances are measured for some fraction of galaxies ($\sim 40\%$ of normal galaxies within 3,000 \kms\ decreasing to only a few percent at 10,000 \kms) and this information can be translated to give maps of peculiar velocities - the radial components of deviations from the cosmic expansion.  To first order in the case of distance measurements, errors with a given methodology are a fixed fraction of distance.  In metric units errors in distance and, hence, peculiar velocities are large at large distance.

For these various reasons our reasonably nuanced view of our immediate neighborhood is not sustained at larger distances.  A fortunate aspect of peculiar velocities, though, is their closer adherence than densities to linear theory \citep{1980lssu.book.....P} over a wider range of environments (Doumler 2012).  Coherent galaxy streaming carries information about structure on scales larger than the domain of distance observations \citep{1995ApJ...449..446Z}.

We will be considering data at various levels of interpretation.  The most basic, but also the most available, come from redshift surveys.  We can know the 3D positions in redshift space of tens of thousands to millions of galaxies.  At a first level of complexity, we can smooth over individual points compensating for the loss of faint galaxies and kinematic stretching in clusters then present results in terms of isodensity contours.  As for peculiar velocities, these measures can be presented raw, or averaged over groups, or otherwised smoothed.  The next step in complexity is to fit models.  Through Wiener Filtering, even with sparse sampling, maps of velocity streaming can be convolved into maps of the most important cosmographic density features.

\section{Mapping Details}

The scales in all the figures are in \kms.  In general galaxies are located in redshift space from knowledge of systemic velocities.  In some instances, especially for plots of the nearest galaxies, galaxies are located by measured positions with the conversion to a velocity scale assuming a Hubble constant
H$_{0} = 74$ \kmsMpc. This conversion is consistent with the relationship between distances and velocities in the data set used in this preliminary study, the data set presented in \citet{2008ApJ...676..184T} and that is now being called Cosmicflows-1.  This conversion is in reasonable agreement with what is being found with an extended data set under development called Cosmicflows-2. With the new compilation we find a slightly higher value of $76$ \kmsMpc\ when calibrating the Tully-Fisher I band slope with nearby SNIa \citep{2012ApJ...749..174C} and $75$ \kmsMpc\ with a similar calibration in the NIR at 3.6 $\mu$m \citep{2012ApJ...758L..12S}.

The supergalactic system of coordinates used in all figures
has been defined by \citet{1991rc3..book.....D}. It is a spherical-coordinate system SGL, SGB
with the North Pole (SGB=$90\deg$) in the direction of Galactic coordinates (l =$47.37\deg$;	b= $+6.32\deg$). 
Cartesian coordinates SGX, SGY, SGZ are derived whenever the recession velocity is known.

Catalogs used in this article are stored in EDD, the Extragalactic Distance Database \citep{2009AJ....138..323T}\footnote{http://edd.ifa.hawaii.edu}.
The names of structures identified in the maps are in part drawn from the HyperLeda Lyon Extragalactic Database \citep{2003A&A...412...45P} and the NASA/IPAC Extragalactic Database (NED).  Maps within 3000 \kms\ were compared with those of the Nearby Galaxies Atlas \citep{1987nga..book.....T}.

The figures presented in this paper and the accompanying movie were created using SDvision, an interactive visualization tool developed within the framework of IDL Object Graphics~\citep{2008ASPC..385..327P}.  This software was initially developed for the three-dimensional exploration of massively parallel numerical simulations such as the Horizon Galaxy Formation simulation~\citep{2009ASPC..406..317P}.  It is being extended now to be used to visualize
redshift surveys by techniques that will be demonstrated~\citep{2011IAUS..277..154P}.

\section{The Movie Begins: The V8k Redshift Dataset} 

Our reference redshift survey is a compilation limited to a cube in supergalactic coordinates with velocities less than 8,000~\kms\ on the cardinal axes.  This compilation of 30,124 galaxies is called `V8k' and it is one of the catalogs available at EDD, the Extragalactic Distance Database.  V8k is an augmentation of John Huchra's `ZCAT', circa 2002\footnote{https//www.cfa.harvard.edu/$\sim$dfabricant/huchra/zcat/}, including the IRAS Point Source Catalog redshift survey and its extension toward the Galactic plane \citep{2000ASPC..218..153S,2000MNRAS.317...55S} and new material particularly relating to nearby dwarfs and HI detections.  The catalog does not include material from recent large but directional redshift surveys.  The density of local coverage compares favorably with the most recent all-sky 2MASS redshift surveys \citep{2005IAUS..216..170H, 2012ApJS..199...26H}.  Whereas essentially every galaxy more luminous than $M_B = -16$ is included at 1000~\kms, only 1 in 13 such galaxies are included at 8000~\kms.  The brighter galaxies are retained, so the completion in luminosity is 40\% at 8000~\kms.  Cluster `finger of god' kinematic artifacts are truncated in V8k (galaxies in known clusters or clusters identified by the kinematic signature of velocity stretching in the line-of-sight are placed at the mean velocity of the cluster, plus/minus 10\% of the galaxy's velocity deviation from the mean).   Distances to galaxies within 3,000~\kms\ are modified to conform to a flow model influenced by the Virgo Cluster \citep{1995ApJ...454...15S}.  The loss of galaxies from the sample was evaluated by fitting a \citet{1976ApJ...203..297S} luminosity function with fixed bright cutoff and faint slope to the observed galaxies in redshift shells, determining the missing contributions from faint end departures from the luminosity function fit, and constraining a polynomial expression to the growth of the missing numbers and flux with redshift.   
The catalog V8k was developed for pedagogic use, for example, in planetarium programing.  It was not intended for quantitative work but, in its defense, V8k provides reasonably uniform coverage around the entire sky except in the zone of obscuration and is more inclusive of the broad range of galaxy types than alternative compilations.  The high density of coverage that it provides suits the qualitative purposes of this paper.

\begin{figure}
%\end{figure}

%\clearpage
%Fig1
%\includegraphics[width=\textwidth]{FIGURES/SDvision_v8k_cosmography_allskymap_v006_dust-W2B-bot45.eps}
\includegraphics[width=\textwidth]{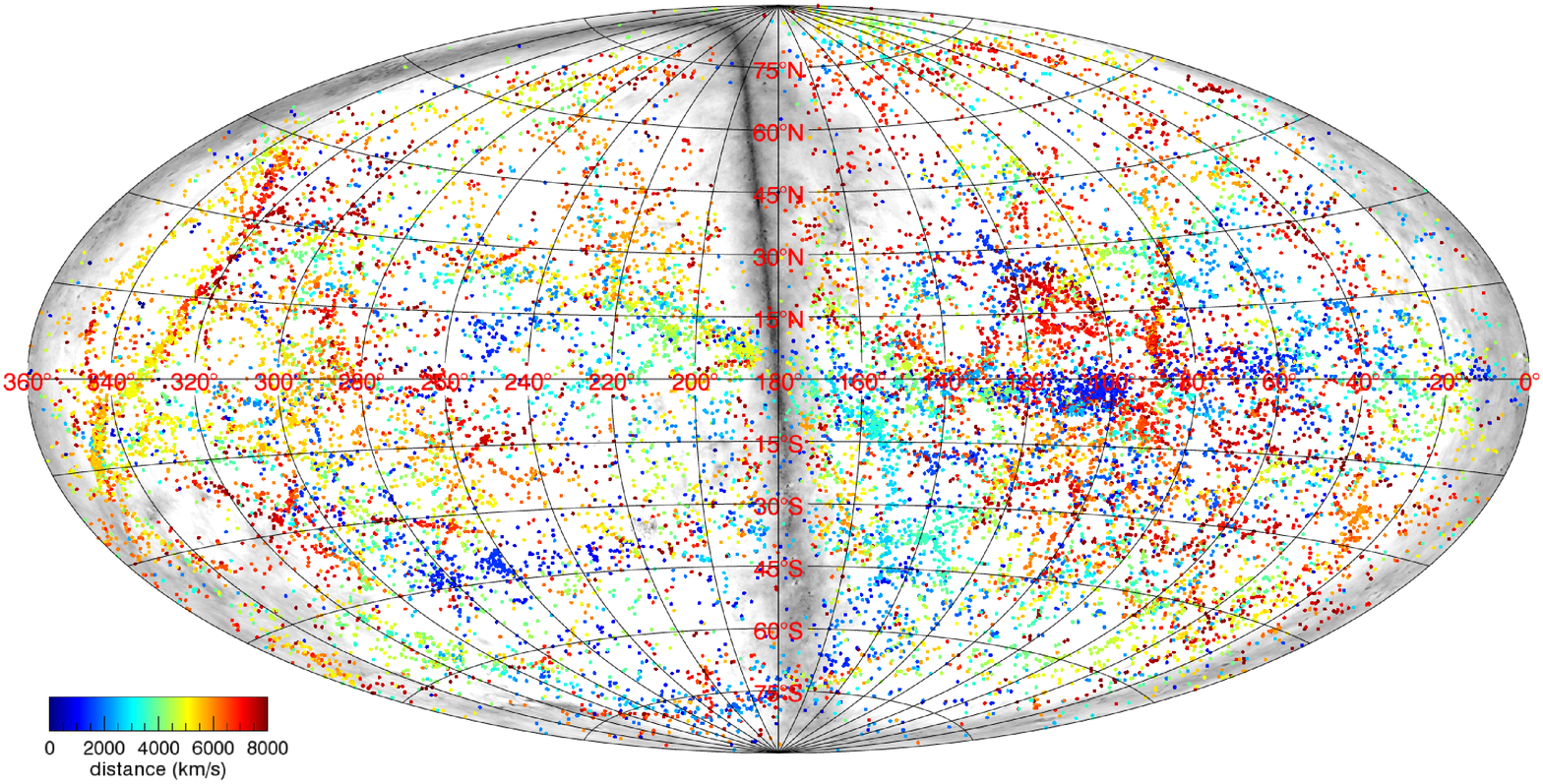}
\caption{Aitoff All-Sky map of the V8k catalog in the supergalactic coordinate system,
restricted to galaxies within $\leq$ 8000 km/s.}
\label{V8K_allsky}

\hspace{1.5cm}

%Fig2
%\begin{figure}
%\includegraphics[width=\textwidth]{FIGURES/SDvision_v8k_cosmography_perspective_v001.eps}
\includegraphics[width=\textwidth]{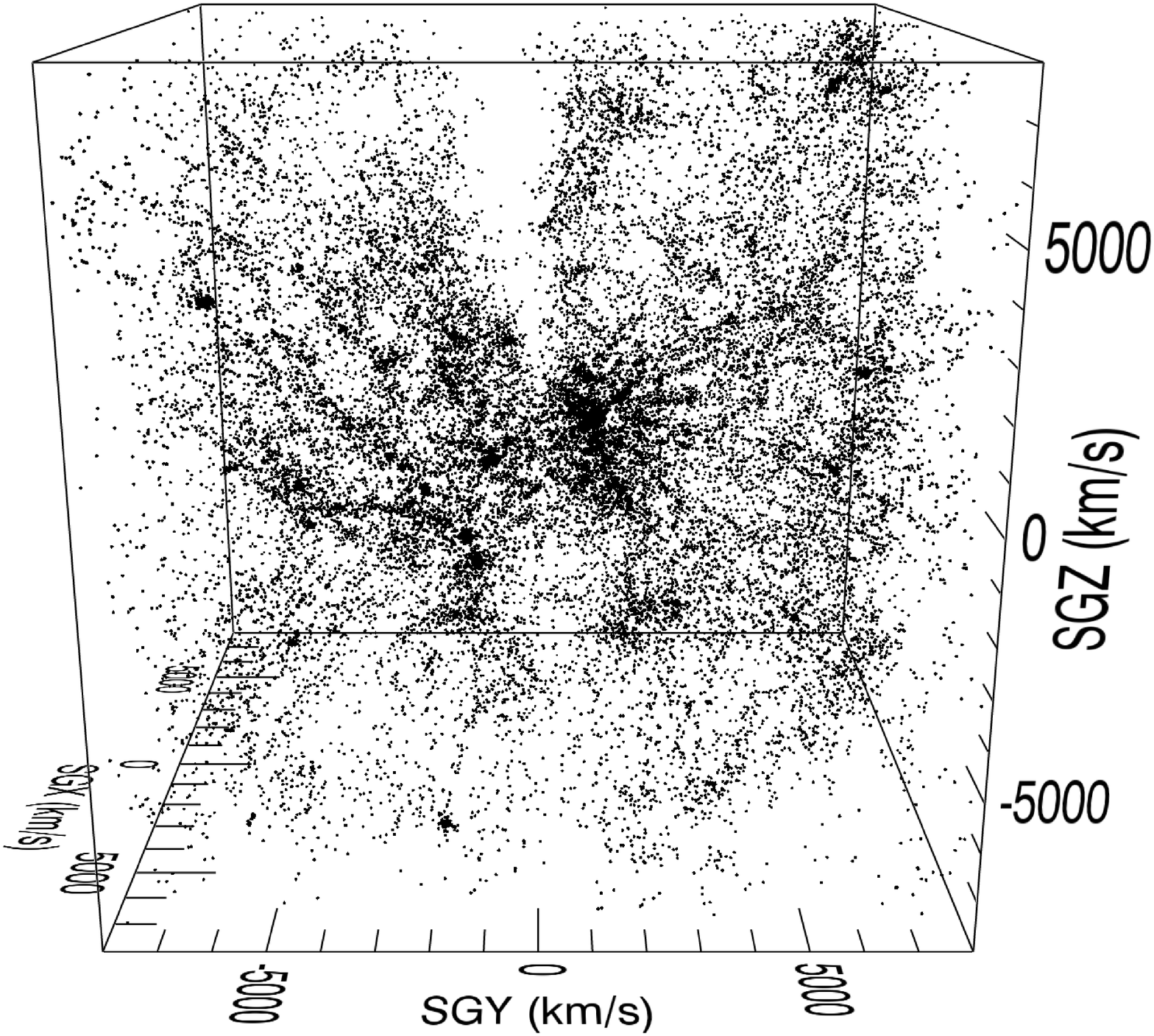}
\caption{A three-dimensional view of V8k redshift catalog. Cluster fingers of god are suppressed.
V8k contains many faint, low surface brightness galaxies not included in alternative compilations like 2MASS.
The zone of avoidance causes the vertical gap and galaxies in the region around the Virgo Cluster give rise to the greatest crowding.}  
\label{V8K}
\end{figure}

We begin with Figures~\ref{V8K_allsky} and \ref{V8K} by showing in the second of these plots a projection of the V8k catalog and in the first an Aitoff projection of those galaxies in the cube that are within an 8,000~\kms\ radius sphere (so loosing the corners of the cube in the Aitoff projection).   The maps are in supergalactic coordinates which, with the Aitoff projection, has the benefits that Galactic obscuration is mostly at the edges except for the central almost vertical crossing and most of the dominant extragalactic structures lie near the equator.  The grey band derives from far-infrared dust emission maps \citep{1998ApJ...500..525S}.

The use of colors to indicate distance in the Aitoff plot helps reveal most of the most important structural elements in the 8,000~\kms\ volume.  There is a great preponderance of nearby galaxies (blue colors) in the right hemisphere (Galactic north) with a node at the Virgo Cluster (SGL=102, SGB=$-2$).  The only significant concentration of nearby galaxies in the other hemisphere is in the vicinity of the Fornax Cluster (SGL=263, SGB=$-42$) and the adjacent Eridanus cloud.  The swath of light blue galaxies in the interval $130 < SGL < 160$ and $-50 < SGB < 0$ lie within the region of the Centaurus and Hydra clusters, the core of the supercluster complex that contains our galaxy at the periphery, and what has been called the `Great Attractor' \citep{1987ApJ...313L..37D}.   A long filament in greens and yellows stretches from this region to the left, to well beyond the Galactic plane in the Galactic south.  This is the Pavo$-$Indus structure \citep{1998lssu.conf.....F}.  Over at the left edge of the Aitoff plot, at a steep angle to the supergalactic equator in yellows and reds, lies the very prominent Perseus$-$Pisces filament \citep{1993AJ....105.1251W}.  Returning to the northern Galactic hemisphere at the right, it is seen that there are many galaxies in red over a wide area behind the Virgo Cluster.  These are objects in the Great Wall \citep{1986ApJ...302L...1D}.  They scatter in projection but they are confined in depth.  The most important part of this structure is in Hercules which falls outside the 8,000~\kms\ sphere illustrated in the Aitoff projection although it is contained within the V8k cube.  The other noteworthy feature of Figure~\ref{V8K_allsky} is the {\it absence} of nearby (blue or green) galaxies over a large area centered around SGL=190, SGB=+45, the region of the Local Void \citep{1987nga..book.....T}.

Figure~\ref{V8K} shows a perspective view of the V8k catalog from an observer location slightly north (positive SGZ) of directly along the positive SGX axis.  The Virgo Cluster within the traditional Local Supercluster  \citep{1956VA......2.1584D}  near the center of the figure appears very prominent. It will be shown later that other structures that do not appear in V8k as prominently play
significant roles in the density field. In this view almost face-on to the plane SGY-SGZ, the vertical zone with no galaxies
is the ``Zone of Avoidance" (ZOA). With the V8k catalog, the Galaxy obscures about 20\% of the
volume of the 8,000 \kms\ cube.

\begin{figure}
\begin{center}
\includegraphics[width=0.67\textwidth]{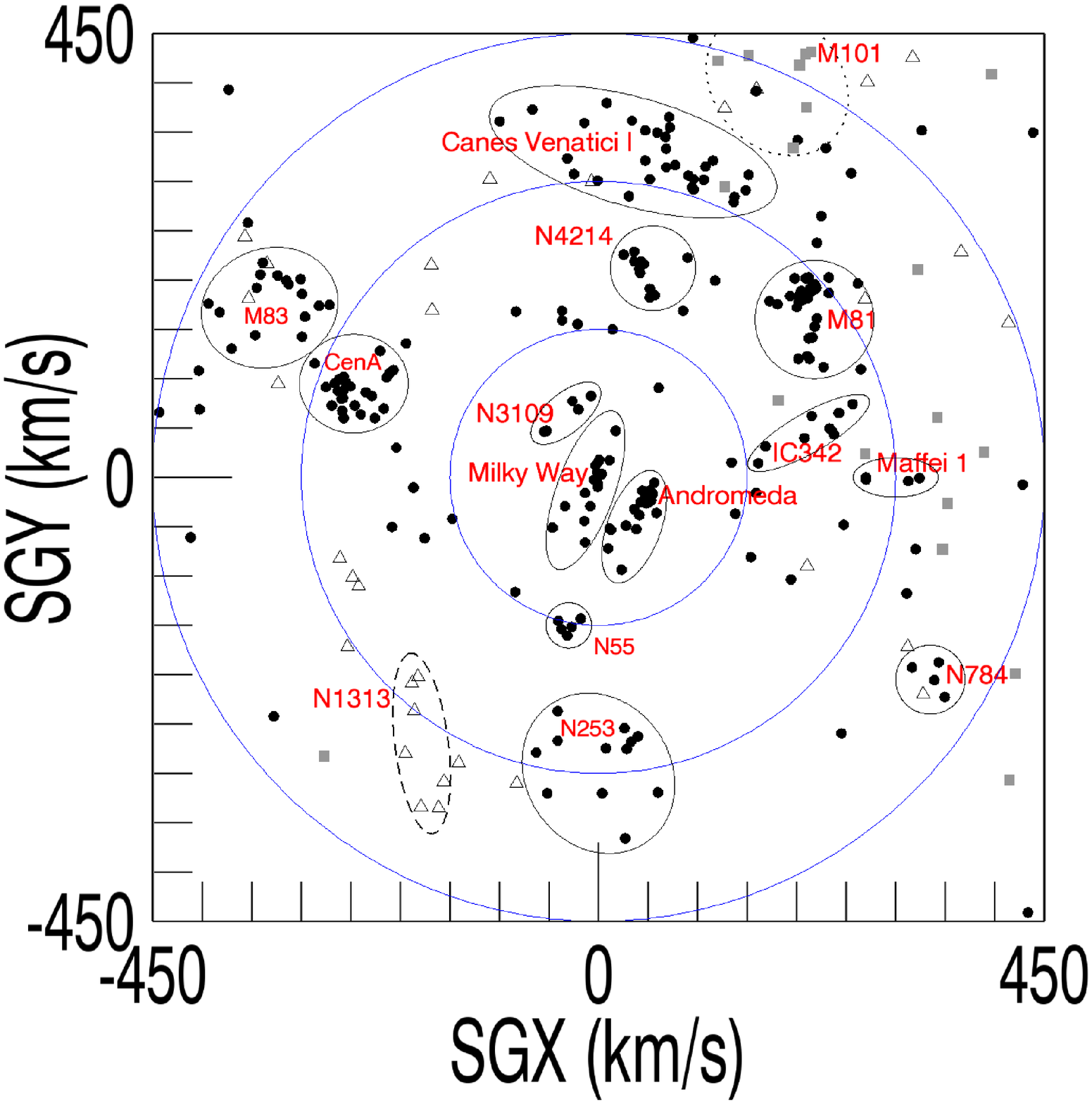}
\includegraphics[width=0.67\textwidth]{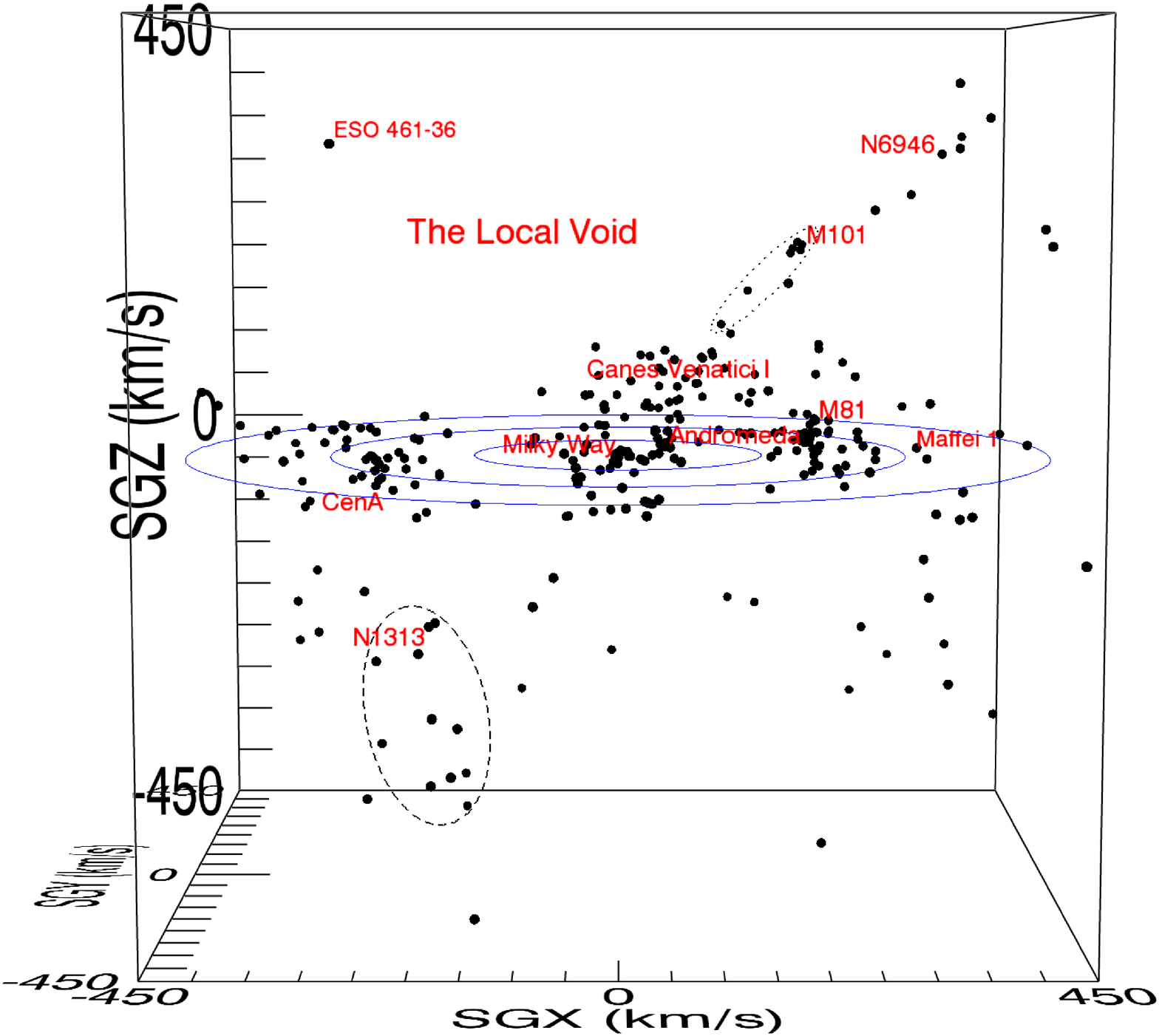}
\end{center}
\caption{Two projections of the Local Group and its immediate vicinity. Three concentric blue circles separated by 2 Mpc = 150 \kms\ lie in the supergalactic plane.  The top panel is a view normal to the plane and gives best separation of the principal groups.  Solid circles, grey squares, and open triangles identify galaxies respectively in, above and below the plane. The view in the bottom panel is almost edge on to the supergalactic plane and emphasizes the concentration of galaxies in this plane and the extreme emptiness of a large region above the plane.}
%\label{ALL75-LOCALGROUPVICINITY}
\label{LG}
\end{figure}

We begin our exploration with a zoom toward the center of the data shown in Figure~\ref{V8K} and a rotation so we are looking down from the positive SGZ axis.  We arrive at the scene of the immediate vicinity of the Local Group shown in the top panel of Figure~\ref{LG}.   The Local Group itself is a region that has separated from the cosmic expansion and is collapsing toward either our Milky Way Galaxy or Andromeda (M31).  The distinct haloes around each of these two major galaxies are destined to merge \citep{2012ApJ...753....9V}.  Other well known nearby groups of galaxies are roughly enclosed by grey ellipses.   The nearest of the substantial groups, the IC342/Maffei complex, lies in the zone of obscuration and is unfortunately poorly studied \citep{1995AJ....110.1584K}.  The complex in Centaurus is found to separate into components around radio galaxy Cen A and, a Mpc further away, M83 \citep{2002A&A...385...21K}.  In addition to the familiar groups around big galaxies, there are associations of dwarf galaxies \citep{2006AJ....132..729T}, suspected to be bound though not relaxed.  The brightest galaxy in the nearest of these is NGC 3109.  This association lies just beyond the Local Group infall domain at 1.3 Mpc.   The groupings around NGC 55 and NGC 4214 are two more nearby dwarf associations.

The view in the bottom panel of Figure~\ref{LG} involves a rotation to a viewing position slightly raised in SGZ from along the negative SGY axis.  The two noteworthy points are the concentration of galaxies to the supergalactic equator, SGZ=0, and the emptiness of most of the volume at positive SGZ, a region that falls within the Local Void.  The one lonely galaxy seen in this space is ESO 461-36  \citep{2008ApJ...676..184T, 2011AJ....141..204K}.

\begin{figure}
\begin{center}
\includegraphics[width=0.85\textwidth]{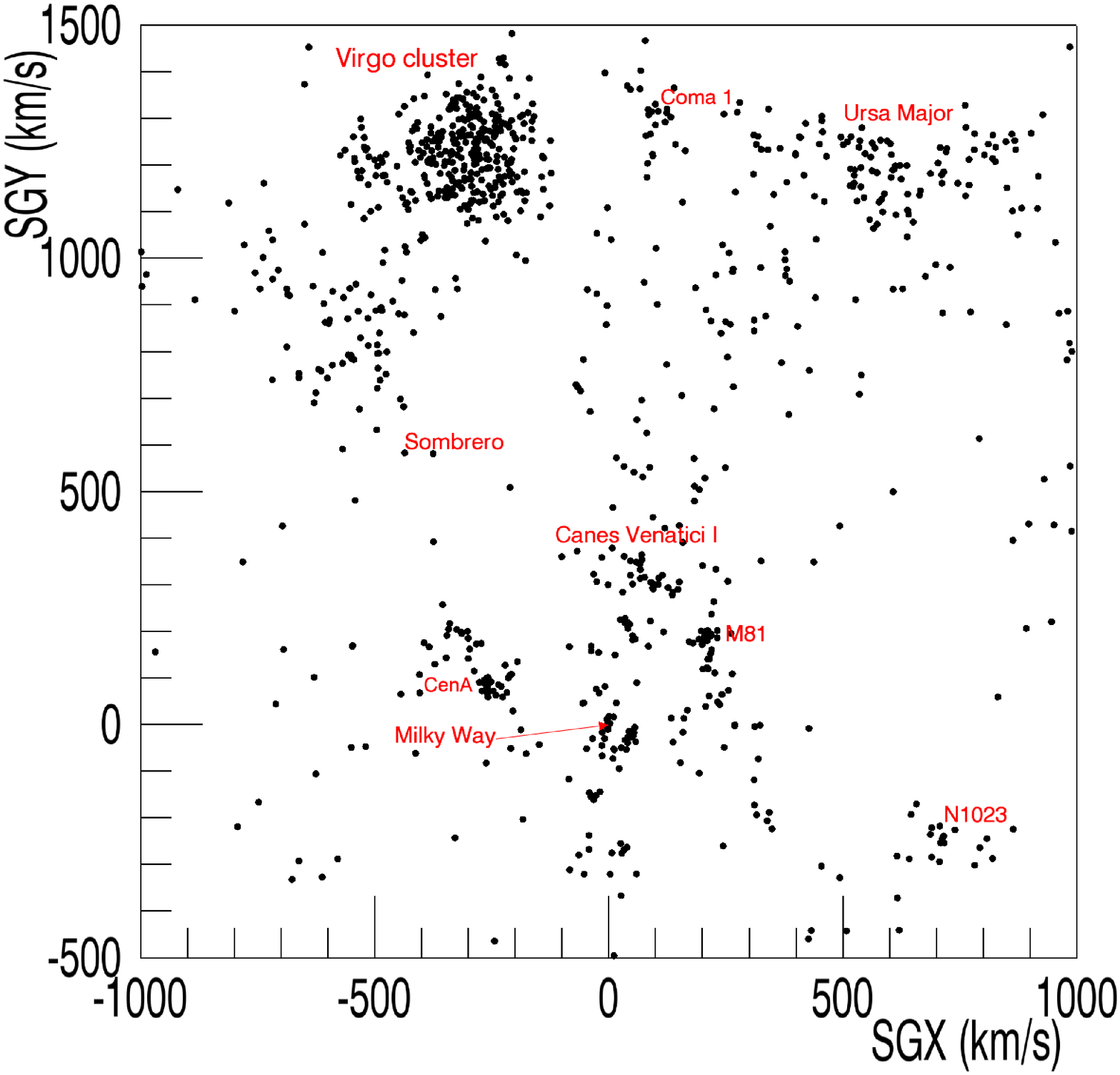}
\end{center}
\caption{A projection normal to the supergalactic plane of a region extending to include the Virgo Cluster.  This slice is narrowly confined to the supergalactic equator with -150 $\leq$ SGZ (km/s) $\leq$ 150.  Distances are provided by many direct measurements in this volume.}
\label{ALL75-SGX-SGY}
\end{figure}

With Figure~\ref{ALL75-SGX-SGY} we have rotated back to the orientation of the top panel in Figure~\ref{LG}, then panned to extend the view to include the Virgo Cluster.  We are showing a very thin equatorial slice in supergalactic coordinates, $-150 <$ SGZ $< 150$ \kms\ ($\pm 2$ Mpc).  We live in a minor spur that we are now calling the Local Sheet  \citep{2008ApJ...676..184T}.  In the direction away from the Virgo Cluster the  spur takes a dogleg over to the NGC~1023 Group \citep{2009MNRAS.398..722T}.  A very tenuous filament continues from the NGC~1023 region, flirting with the Galactic plane, to reach all the way to the Perseus$-$Pisces chain.

\begin{figure}
\includegraphics[width=\textwidth]{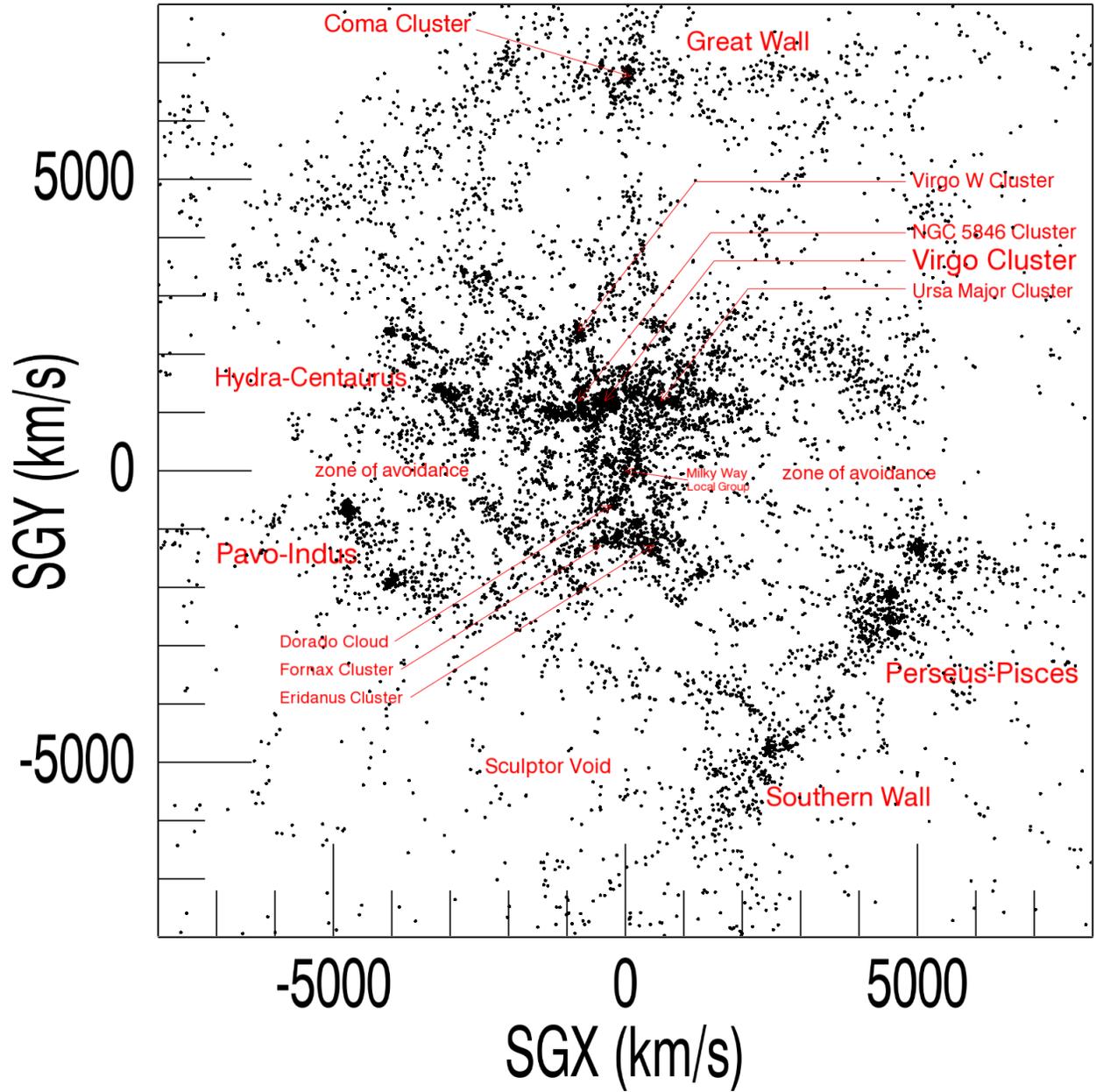}
\caption{A projection normal to the supergalactic plane extending across the full V8k catalog.  The view is limited in depth to a slice with -1000 \kms\ $\leq$ SGZ  $\leq$ 1000 \kms.}
\label{V8K-SGX-SGY}
\end{figure}

To continue the exploration of the V8k catalog, we view from the same direction but pan outward to display a slice of $\pm$ 1,000 \kms\ thickness centered on the SGZ=0 supergalactic plane in Figure~\ref{V8K-SGX-SGY}. The zone of avoidance is horizontal at the mid-plane of the figure and the Virgo Cluster is slightly above the center. Virgo is a member of a chain of clusters that start at the right near the Ursa Major Cluster and is elongated along the SGX axis towards the region labeled Hydra$-$Centaurus.    A prominent structure on the other side of the zone of avoidance is labelled Pavo$-$Indus.  \citet{1998lssu.conf.....F} links this structure to structure across the zone of avoidance and calls the elongated entity the Centaurus Wall.  This connection is probably valid but due to ambiguities in details we retain separate identities.
Also just below the zone of avoidance are the Fornax Cluster and associated Eridanus and Dorado complexes, part of a structure historically called the Southern Supercluster \citep{1953AJ.....58...30D}.  The remaining dominant features in this map are, at negative SGY,  the Perseus$-$Pisces chain \citep{1993AJ....105.1251W} linked to the Southern Wall \citep{1994ApJ...424L...1D} and, at positive SGY, the Great Wall containing the Coma Cluster \citep{1986ApJ...302L...1D}.  Both these major features at the edges of this figure are steeply inclined to this projection.  Finally in connection with this plot, it is worth noting the Sculptor Void, an empty region foreground of the Southern Wall \citep{1990AJ.....99..751P, 1998lssu.conf.....F}.

\begin{figure}
\includegraphics[width=\textwidth]{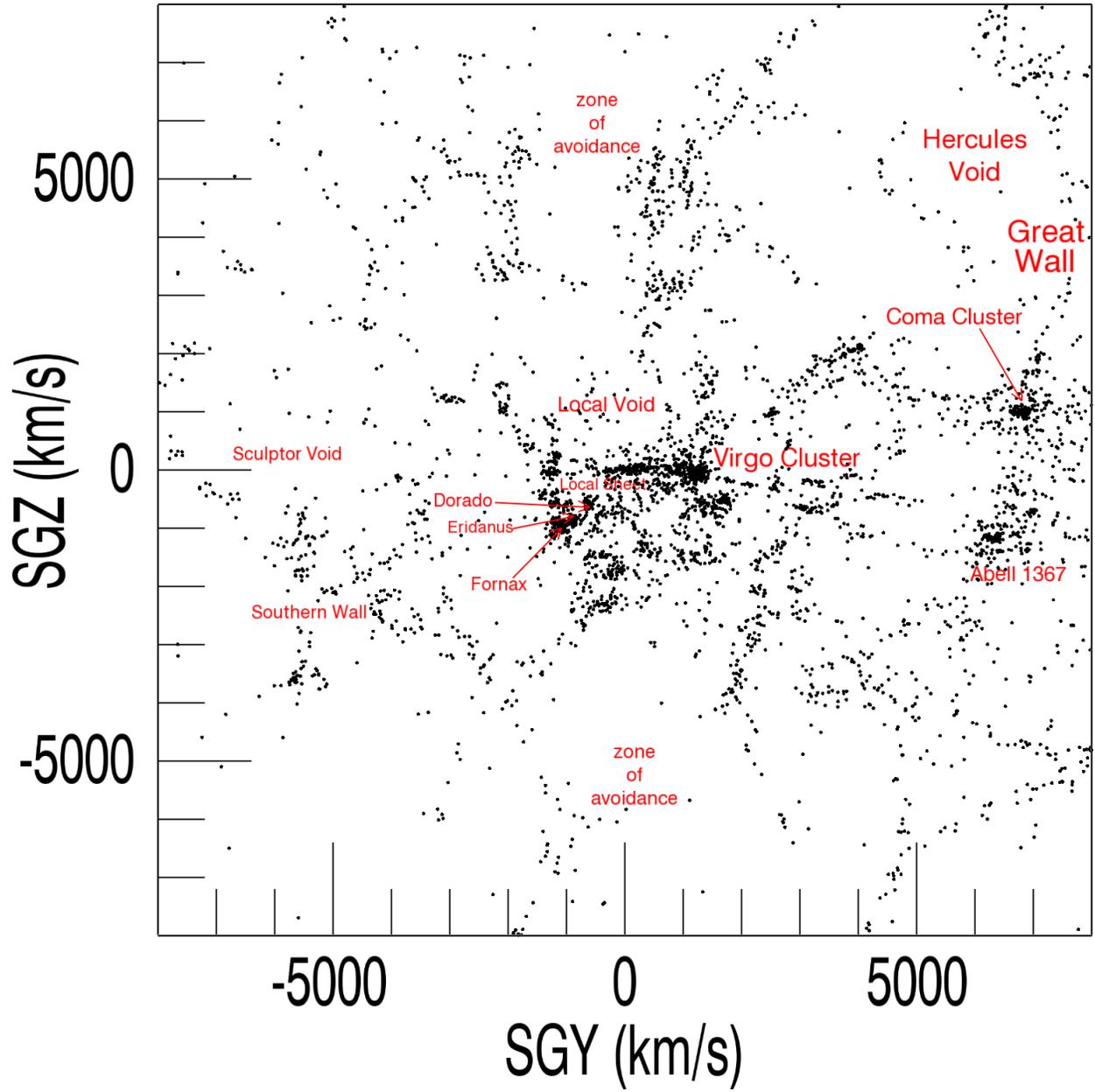}
\caption{An orthogonal projection with the supergalactic equator SGZ=0 at the mid-plane.  The depth is bounded by -500 \kms\ $\leq$ SGX $\leq$ 500 \kms.}
\label{V8K-SGY-SGZ}
\end{figure}

We will now look at the distribution of the V8k sample perpendicular to the supercluster plane with Figure~\ref{V8K-SGY-SGZ}, in an SGY-SGZ projection within a thin slice of $\pm$ 500 km/s about SGX=0.   The view captures some of the Great Wall that runs vertically up the right edge of the map through the Abell 1367 and Coma clusters.  This figure illustrates the filamentary nature of galaxy networks and the prevalence of voids.  For example there is the prominent Hercules Void in the upper right corner of the map centered in the constellation of Corona Borealis that lies to the foreground of the Great Wall and extends in depth a further 2000~\kms\ in negative SGX to the important Hercules Cluster.  Also a filament connecting the Coma Cluster to the vicinity of the Virgo Cluster is to be noted \citep{2008AJ....135.1488T}.
Cosmological simulations, and in general the theory of structure formation and hierarchical scenario of galaxy evolution, have predicted that big clusters of galaxies like Coma should be at a node of filaments. 

%The void which seems to be enclosed by Coma and Hercules supercluster and their filaments, has an average dimension of about 6,000km/s or 80 Mpc of diameter. \textcolor{red} {HC:  I propose to name this new void the Hercules Void}. This is also the size of the largest dimension of the Local Void being enclosed on one side by our local supercluster. 

Coming back to Virgo and the Local Supercluster at the center of the map, it is seen that the nearest galaxies lie in a flat, thin structure.
This ``Local Sheet" is the nearby region where the
Milky Way, the members of the Local Group and other galaxies share similar peculiar velocities.
The Local Sheet is part of the wall bounding the Local Void, seen at SGZ positive and $-20 < {\rm SGY} < +10$ Mpc \citep{1987nga..book.....T, 2008ApJ...676..184T}.   

%There is also a clearly seen void directly behind the Virgo Cluster at SGY $\sim +30$ Mpc and SGX $\sim 0$ that we will call the `Virgo Void'. The Local and Virgo voids are linked by a generally under dense region above and under the Local Supercluster at SGZ $\sim 15$ Mpc. 

\begin{figure}
\includegraphics[width=\textwidth]{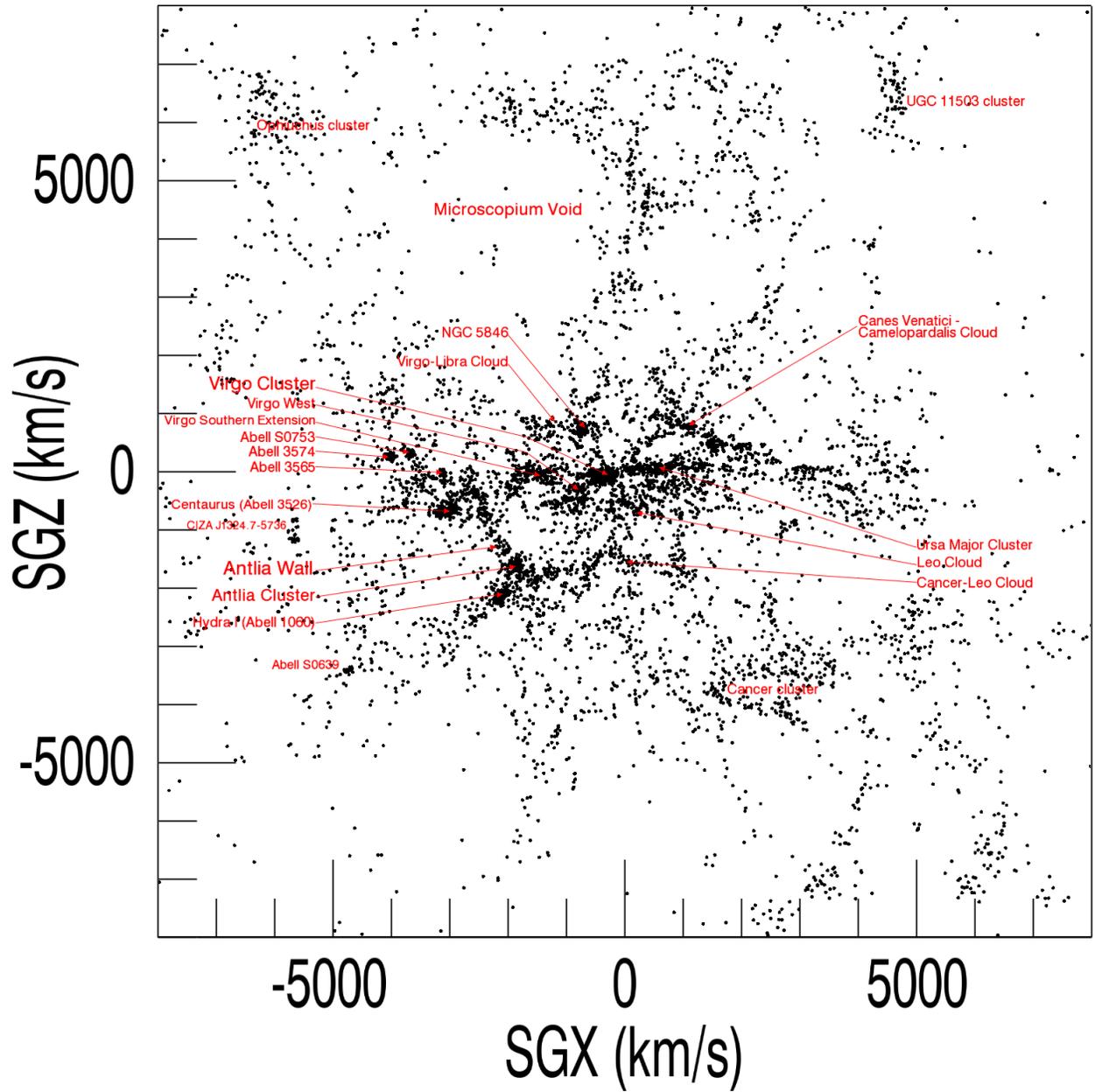}
\caption{The other orthogonal projection, again with the supergalactic equator SGZ=0 at the mid-plane.  In this case the slice is displaced from the origin to 500 \kms\ $\leq$ SGY $\leq$ 2500 \kms\ to center on the main components of the Local and the Hydra-Centaurus supercluster complexes.} 
\label{V8K-SGX-SGZ}
\end{figure}

Finally in this group of figures, we rotate in Figure~\ref{V8K-SGX-SGZ} to the third cardinal axis to show galaxies in the SGX-SGZ plane that lie in the range 500 \kms\ $\leq$ SGY $\leq$ 2500 \kms. This slice is chosen to isolate most of the main components of our home supercluster complex.  The historical Local Supercluster considered by \citet{1956VA......2.1584D} only included the region within $\sim 2000$~\kms\ of the Virgo Cluster dominated by the concentration of galaxies to SGZ $\sim 0$.  This filament extending horizontally from Virgo was already seen in Figure~\ref{V8K-SGX-SGY}.  It connects to the Centaurus Cluster, A3526, which proves to be a nodal point for the confluence of five important filaments.  In addition to the filament running horizontally to the right to the Virgo Cluster, there are spokes running down and to the right through the Antlia Cluster (S636), up and to the left through four Abell clusters (A3574, A3565, A3537, and S753), and two spokes out of the page to negative SGY, one through the Norma Cluster (A3627) then along the Pavo$-$Indus arm, and another reaching in to the Fornax Cluster.  The filament down and to the right toward the Antlia Cluster is particularly prominent in Figure~\ref{V8K-SGX-SGZ}.   \citet{1998lssu.conf.....F} refers to this feature as the Hydra Wall although the Hydra Cluster (A1060) lies to the background of the structure.  It has long been suspected that this region including the Centaurus, Hydra, and Norma clusters is strongly influencing the flow pattern of galaxies including the Milky Way \citep{1987ApJ...313L..37D}.  This matter will come up again later.  We summarize the current discussion by stressing that de Vaucouleur's Local Supercluster is an appendage of a substantial supercluster centered on the Centaurus Cluster.  

Figure~\ref{V8K-SGX-SGZ}, like the previous two plots, is noteworthy in its display of filaments and voids.  We draw attention to a particularly large void centered at SGX=$-2000$ \kms, SGZ=4500 \kms.  \citet{1998lssu.conf.....F} calls this feature the Microscopium Void and notes that it stretches forward in SGY from the slice we show across the zone of obscuration.

\section{From Individual Galaxies to Iso-density Surfaces}

An evident deficiency with plotting individual galaxies, whether drawn from V8k or any other catalog, is the bias resulting from increasing incompleteness as a function of distance.  Before an evaluation can be made of the relative importance of local structure, an attempt must be made to correct for the loss of information with distance.  An example of such an exercise is the study by \citet{1993MNRAS.265...43H} who generated density contour maps after making corrections for incompletion.   The first step is to evaluate the intrinsic properties of a complete sample of galaxies, say, as characterized by the \citet{1976ApJ...203..297S} luminosity function.  In that early work by Hudson only diameters were available over most of the sky so he had to determine an equivalent diameter function.  The next step is to formulate how deeply into the luminosity (or diameter) function one is sampling as a function of distance.  This analysis gives a measure of how much luminosity is being lost in increasing distance shells.  Here we assume a linear relation between redshift and distance, a reasonable assumption: corrections for lost galaxies increase with distance from negligible to important, but over the same range, relative deviations between redshift and distance decrease from significant to small.  The correction is made in the form of a fifth order polynomial fit to the depletion of galaxies in successive shells in redshift.  The correction is made with an 
adjustment to the luminosities of galaxies in the catalog, effectively an assumption that the missing galaxies reside near the bright galaxies that have been included.  The increase in the amount of lost light is modeled by a fourth order polynomial fit to the amount of lost light from an assumed Schechter function in successively increasing velocity shells.  The resulting sample is smoothed on a grid with a gaussian smoothing of 100 \kms\ within 1000~\kms\ but that increases with distance so the peak compared with an unadjusted luminosity is constant but the luminosity is spread over an increasing volume with distance.  The correction is made to luminosity rather than number through the assumed Schechter function.  The number of lost galaxies becomes very large at the edge of the survey but the lost luminosity is modest since most of the light in a complete sample is in the brightest galaxies which are included.  The adjustment factor at 8,000~\kms, the radial dimension of the data cube on the cardinal axes, is a factor 2.5.  The adjustment factor grows to 10 in the extreme corners of the cube at almost 14,000~\kms.  Blue luminosities are used since that was what was available when V8k was constructed.  
%Greater elaboration of the selection function adjustments seems hardly warranted since V8k is not a rigorously defined catalog and our use here is qualitative.

\begin{figure}
\includegraphics[width=\textwidth]{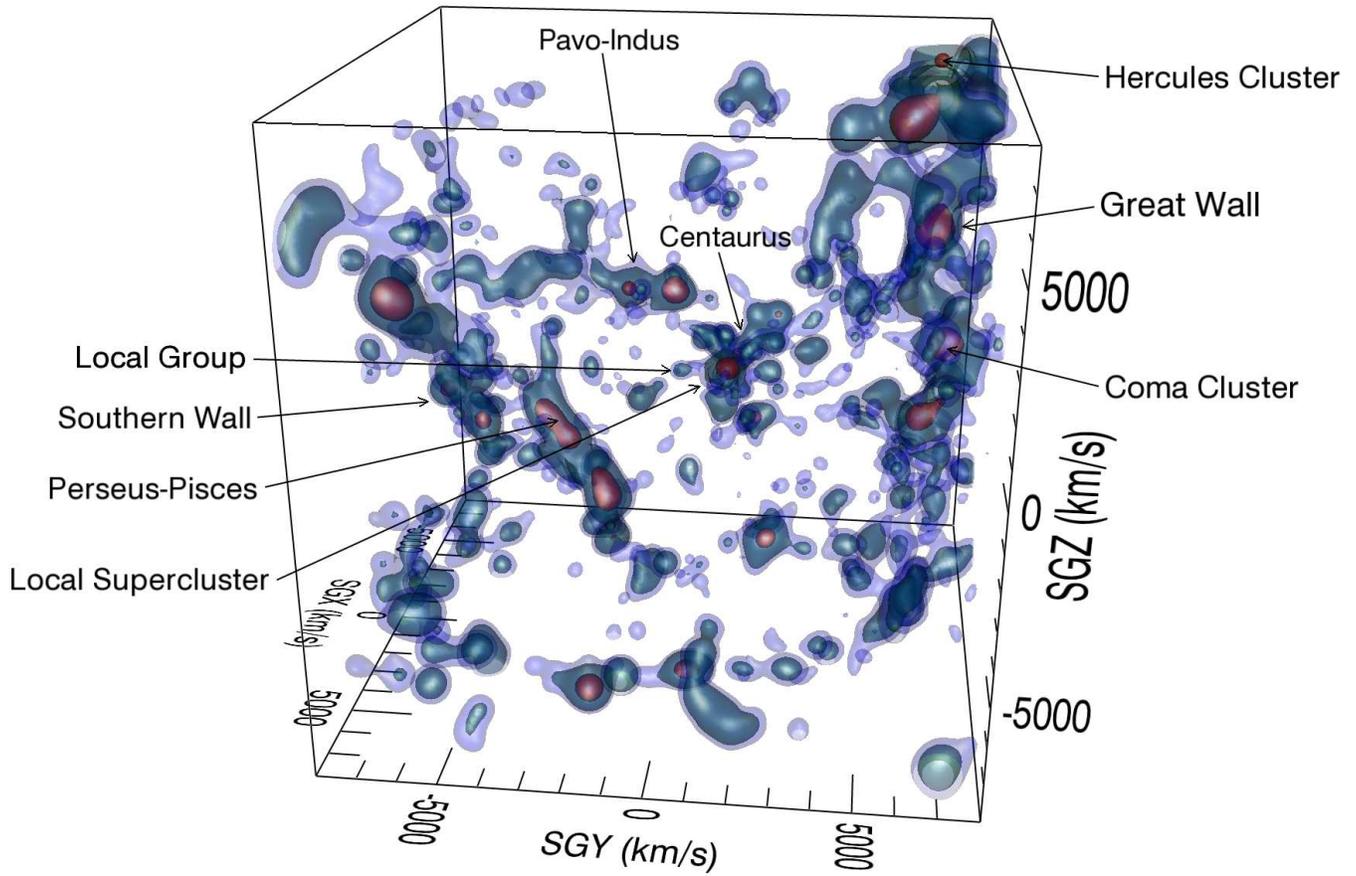}
\caption{A perspective view of the V8k catalog after correction for incompleteness and represented by three layers of isodensity contours.
The region in the vicinity of the Virgo Cluster now appears considerably diminished in importance.  The dominant structures are the Great Wall and the Perseus-Pisces chain, with the Pavo-Indus feature of significance.}
\label{V8K_corr_perspective}
\end{figure}

\begin{figure}
%\begin{center}
%\includegraphics[width=\textwidth]{FIGURES/SDvision_v8k_corr_intensity_cosmography_topview_v005.eps}
\includegraphics[width=\textwidth]{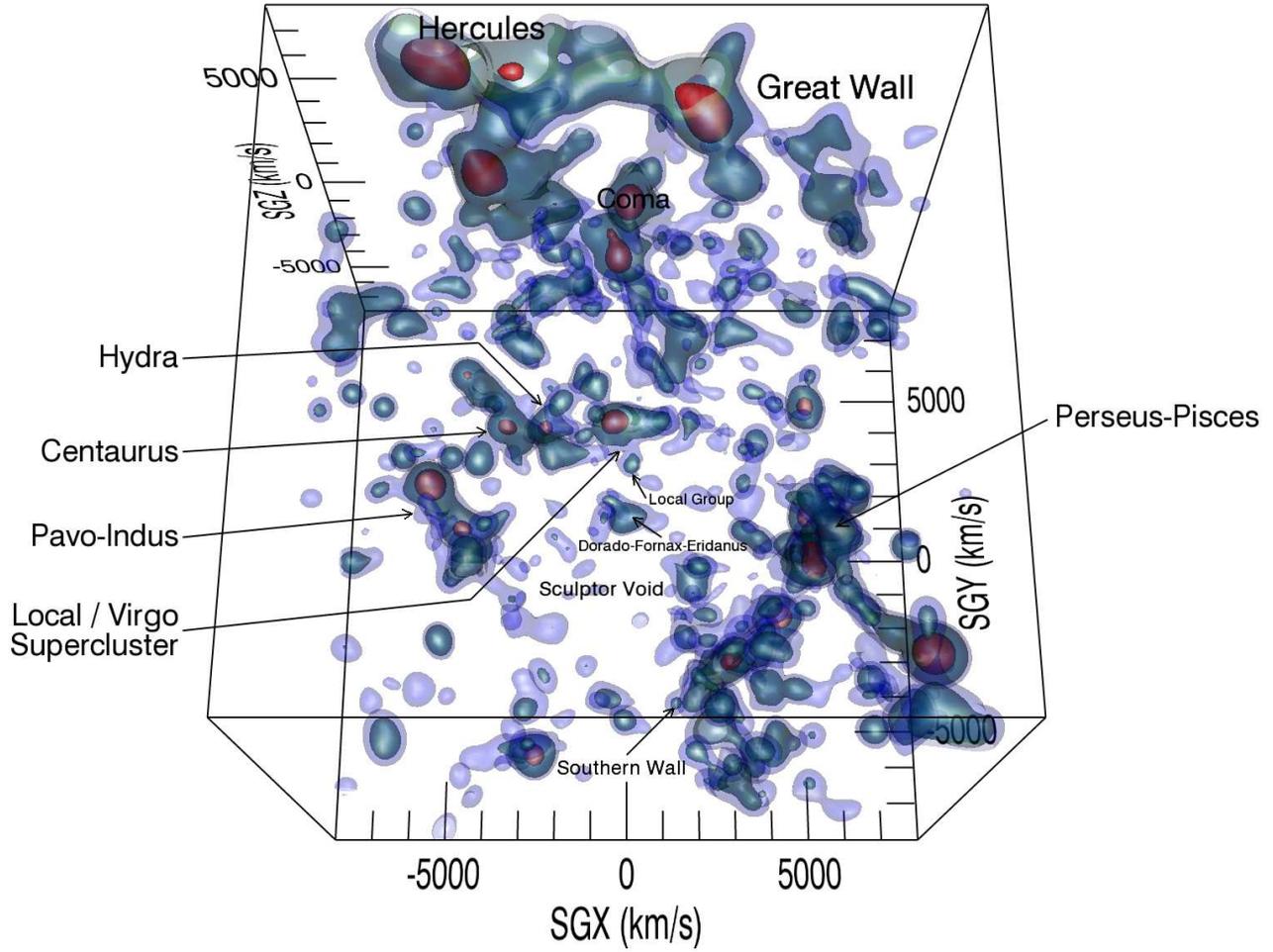}
\caption{A perspective view from almost normal to the supergalactic plane, again corrected for incompleteness and represented by isodensity contours.  The Local Supercluster is seen to be of modest importance.  The Great Wall is dominant in this volume especially in the vicinity of the Hercules Cluster.  The Sculptor (Southern) Wall separates from the Perseus-Pisces chain in this view.} 
%\end{center}
\label{V8K_corr_top}
\end{figure}

In the accompanying movie, the transition from displays of individual galaxies to maps adjusted for incompleteness is shown by a dissolve to a density isosurfaces plot.  We nest cuts at three density levels, at 0.3, 0.1, and 0.07 $L^{\star}$ galaxies per (100~\kms)$^3$ respectively.  Two roughly orthogonal perspective views are given in Figures~ \ref{V8K_corr_perspective} and \ref{V8K_corr_top}.
The isodensity values associated to the surfaces are chosen to display a large range
of structures, from the Local Group to dense clusters and the Great Wall, and so that the major structures appear well defined and separated. 
In Figure~\ref{V8K_corr_perspective} the observing point is roughly edge on to the Great Wall and 
isolates the Virgo/Local Supercluster from its neighbors. The Great Wall appears as a massive superstructure running up to the Hercules Cluster.
It is now clear that what has been called the Virgo or Local Supercluster is a relatively minor element within the V8k volume.  The two dominant structures are the Great Wall and the Perseus$-$Pisces filament.  The Southern (Sculptor) Wall is a significant appendage to Perseus$-$Pisces.  Likewise, the Pavo$-$Indus filament is a substantial, even dominant appendage to the Centaurus$-$Hydra core of the supercluster that we live in.  

In the view from near the supergalactic pole provided by Figure~\ref{V8K_corr_top} we can appreciate the extent of the Great Wall. It covers an entire face of the V8k data cube.  On the opposite side of the galactic plane and at a comparable distance is the Southern Wall.
This wall delimits another giant void, the Sculptor Void with typical dimensions of about 5,000 \kms\ in diameter, quite similar to what is observed
for the Local Void and Hercules Void.
From this viewpoint, we see the connectedness of the Local Supercluster and Pavo$-$Indus features to the Centaurus$-$ Hydra complex.

\section{Centaurus Cluster: A Local Node}
\label{sec:node}

The Centaurus Cluster is known to be close to the direction of large scale flows, the subject of the next section.  Before turning to what can be learned from galaxy motions, there is a point to be made from detailed inspection of the maps that have been presented.    The Centaurus Cluster is special as a focal point for multiple filaments within the neighboring `cosmic web'.  The concept of `filament' is ill defined because there can be large variations in morphology (sheet/string blends, curvature, splitting) and densities or density contrasts. In the present instance two particular difficulties arise in trying to study these structures.  First, incompleteness is becoming a serious problem at distances greater than $\sim 3500$ \kms, not much more than the distance to Centaurus Cluster.  Second, the cluster is near the zone of avoidance.  
In spite of these challenges, using a visualization tool that allows for inspection through rotation and zoom we qualitatively discern five filaments converging at the Centaurus Cluster. 
Four of the five filaments that will be described are affected to some degree by one or other of the above-mentioned problems.  Three cross the ZOA and two radiate to large distances.  Our limited intent with this study is to provide a qualitative look at this nearest major node of multiple filaments using the visualization power of motion.

\begin{figure}
\includegraphics[width=\textwidth]{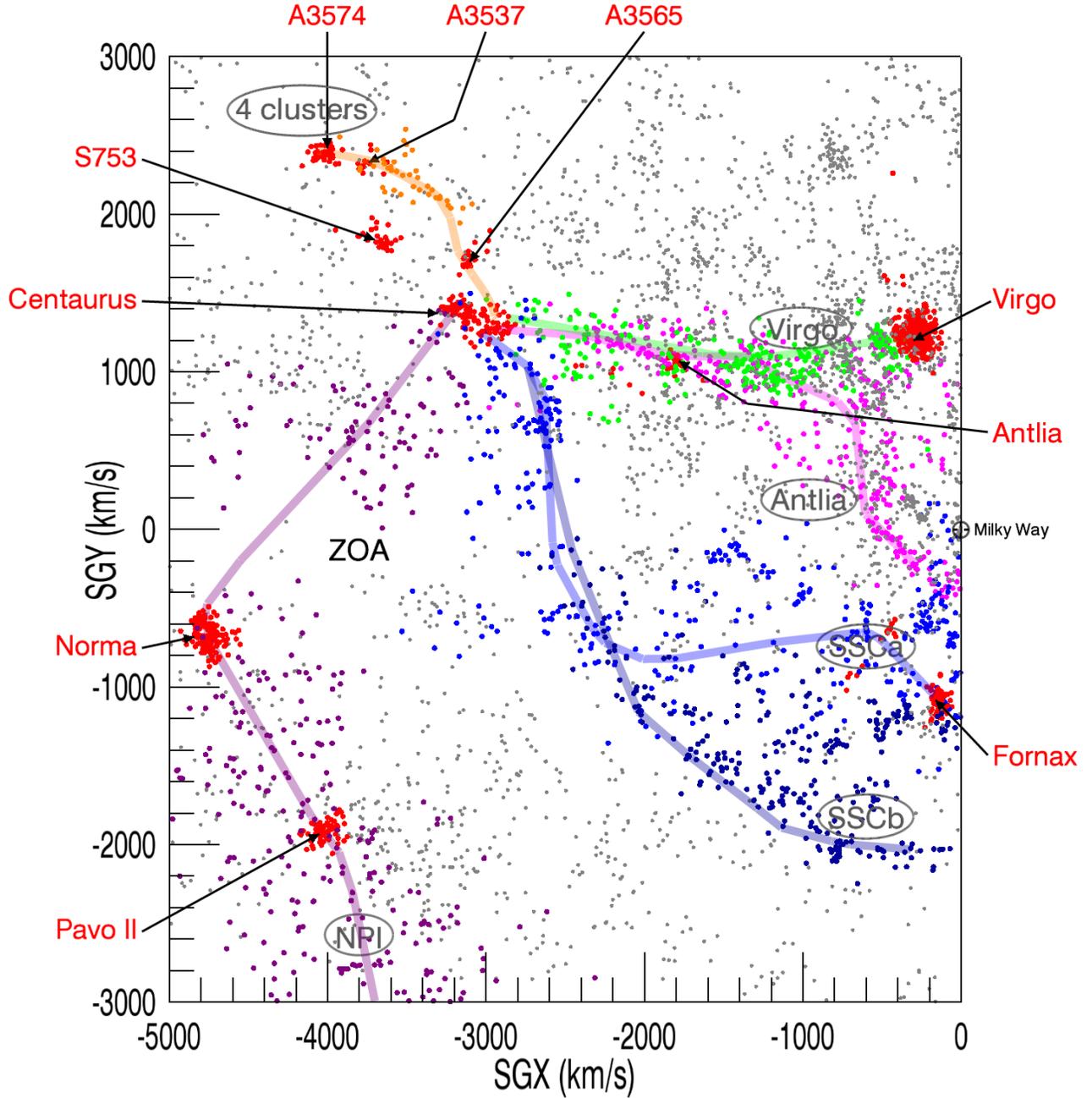}
\caption{Five filaments feeding into Centaurus Cluster: supergalactic pole view.  Major clusters are in red and the 5 filaments identified by names in ovals are given separate colors: Antlia in magenta, Virgo in green, A3537/A3565/A3574/S753 (4 clusters) in orange, Norma-Pavo-Indus (NPI) in purple, and the forking southern supercluster (SSC) in two shades of blue.  The solid bands in the same colors provide outlines.  The cube extends in SGZ over the range $\pm 2000$ \kms.  Galaxies in this box but outside the identified structures are in grey.}
\label{5strands_xy}
\end{figure}

\begin{figure}
\includegraphics[width=\textwidth]{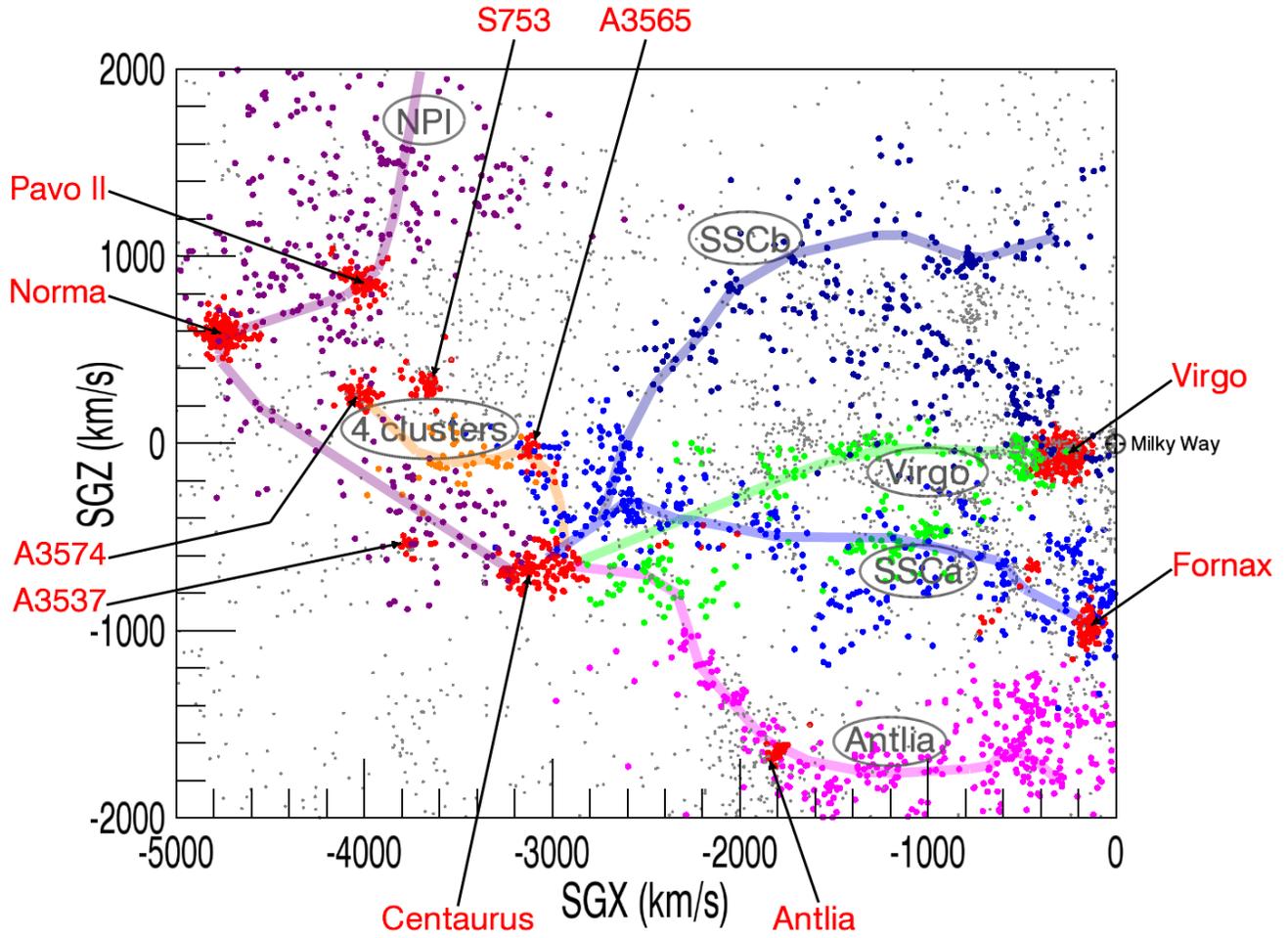}
\caption{Five filaments feeding into Centaurus Cluster: in the supergalactic plane approximately normal to the Milky Way.  Names and color coding as in previous figure.  Same box boundaries.}
\label{5strands_xz}
\end{figure}

\begin{figure}
\includegraphics[width=\textwidth]{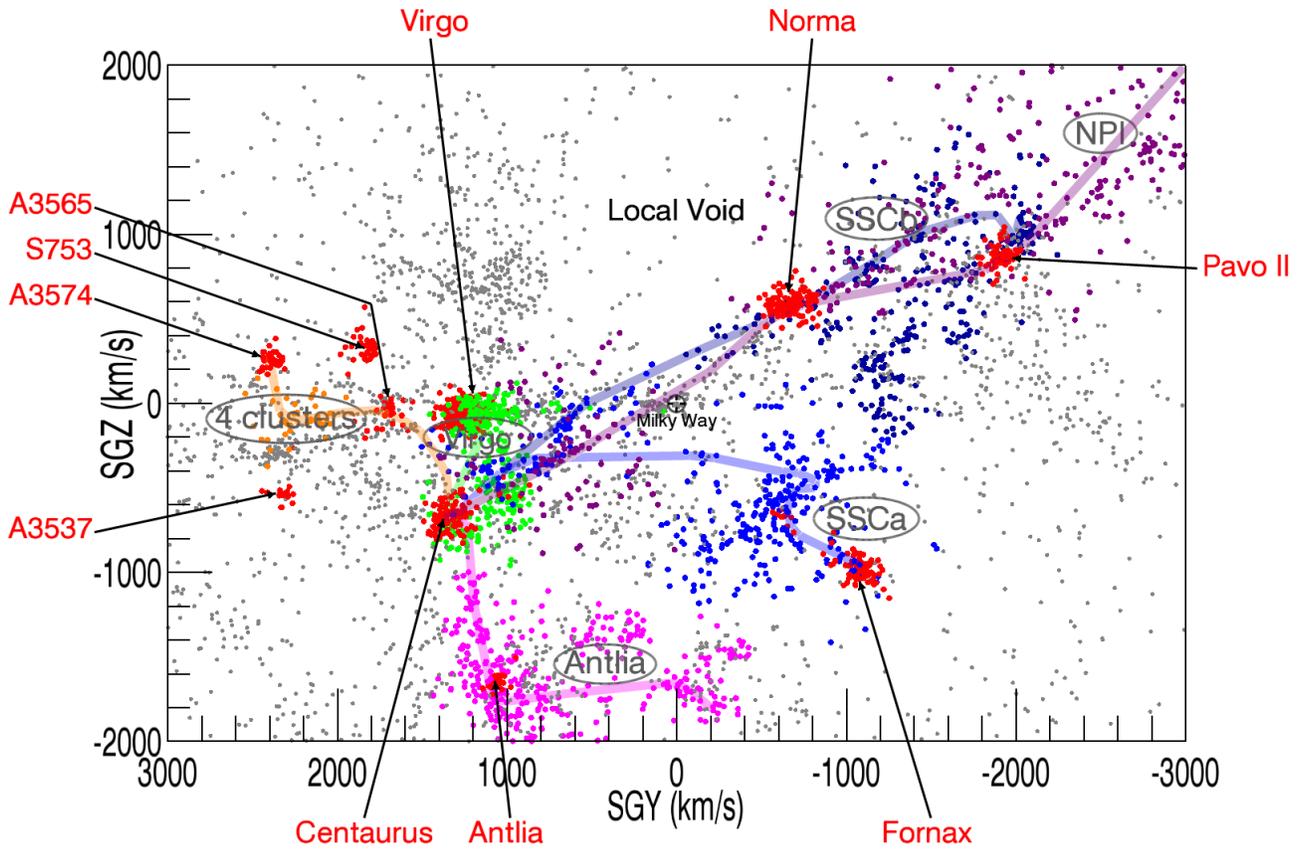}
\caption{Five filaments feeding into Centaurus Cluster: third orthogonal view.  Same box, names, and color codes as previous 2 figures.  Milky Way ZOA is approximately vertical at SGY=0.}
\label{5strands_yz}
\end{figure}

We offer three orthogonal views of the region encompassing most of the filaments emanating from the Centaurus Cluster with Figures \ref{5strands_xy}, \ref{5strands_xz}, and \ref{5strands_yz}.   The case we are trying to make is visually much clearer in the accompanying movie than in the static figures.  However the figures capture the salient points.  In the movie, we begin with only a small rotation from the previous figure to a view from the north pole in supergalactic coordinates and then zoom to a restricted region around the Centaurus Cluster.  There is also a dissolve to individual galaxies and a scheme of colors that will help in a discussion of individual features.  The colors and naming conventions are specified in the caption for Figure~\ref{5strands_xy}, the first of the three frozen scenes.  Galaxies in 11 clusters are colored red, including those in the Centaurus Cluster, our particular interest.  Galaxies in the five filaments of interest are given separate colors.  The structures can be followed in the movie through a tilt to a view looking in from the negative SGY axis (Figure~\ref{5strands_xz}), then a rotation to a view looking in from the negative SGX axis (Figure~\ref{5strands_yz}).

The most familiar strand is the connection between the Centaurus Cluster and the Virgo Cluster because this feature extends in our direction at high Galactic latitudes.  This structure is colored green and labelled `Virgo'.  It extends through the Virgo Cluster and continues through the Ursa Major Cluster, out of the box at positive SGX (see Figure~\ref{V8K-SGX-SGY}).  This structure is the main component of what has been called the Local (Virgo) Supercluster.  In detail it is composed of two strands, one above the other in SGZ (see Figure~\ref{5strands_xz}).  The upper branch has historically been called the Virgo Southern Extension \citep{1956VA......2.1584D} and with the continuation through Ursa Major has provided the main constraint in the definition of the supergalactic equator and supergalactic coordinates.  The lower branch is a lesser structure called the Crater Cloud in the Nearby Galaxies Atlas \citep{1987nga..book.....T}.  It continues out of the box to positive SGX as the structure called the Leo Cloud in the Nearby Galaxies Atlas (identified in Fig.~\ref{V8K-SGX-SGZ}).

A second strand out of the Centaurus Cluster is quite pronounced in Figure~\ref{V8K-SGX-SGZ} where it is labeled Antlia Wall.  In Figures \ref{5strands_xy}, \ref{5strands_xz}, and \ref{5strands_yz} this structure is given the color magenta (except where it merges with the Virgo filament in green) and is labeled `Antlia'.  It was mentioned earlier that \citet{1998lssu.conf.....F} called this feature the Hydra Wall but we noted that the Hydra Cluster is in the background.  This structure has two major inflections.  The first leg joining the Centaurus and Antlia clusters is very prominent.  At the Antlia Cluster the filament takes a turn to run at a roughly constant SGZ $\sim -1700$~\kms\ (see Fig. \ref{5strands_xz}).  Over these first two legs the filament lies almost directly under  (lower SGZ than) the Centaurus-to-Virgo filament at SGY $\sim 1200$ \kms, but at SGX $\sim -800$ \kms\ the structure takes a turn of $\sim 90^{\circ}$ to head under the ZOA (see Fig. \ref{5strands_xy}) to link with the structure called the Lepus Cloud in the Nearby Galaxies Atlas.

A third and very extensive structure linked to the Centaurus Cluster passes through the Norma Cluster and along the Pavo-Indus filament through the Pavo II Cluster.  This structure is labeled NPI and associated galaxies are given the color purple.  \citet{1998lssu.conf.....F} called the structure the Centaurus Wall and it has received a lot of attention in the region of the ZOA because of the proximity in direction with the motion of our Galaxy inferred from the Cosmic Microwave Background dipole  \citep{1996ApJ...473..576F}.   These studies have revealed the importance of the Norma Cluster \citep{1996Natur.379..519K}.  The structure is the most striking feature in Fig. 4 of the review article by \citet{2000A&ARv..10..211K}.  It is easily the longest of the structures directly connected to Centaurus.  It was entertained that this feature might directly connect with the important Perseus-Pisces chain \citep{1996MNRAS.283..367D} through the Southern Wall.  The Pavo-Indus filament, the Southern Supercluster filament (soon to be discussed), and the Southern Wall are sectional boundaries of the immense Sculptor Void \citep{1998lssu.conf.....F}.  In Figure \ref{5strands_xy} the data gap across the ZOA to Norma is large but the linkage is more apparent in dedicated surveys of the region \citep{2006MNRAS.369.1131R}.  The filament extends beyond the display box.

The fourth prong extending from the Centaurus Cluster is probably poorly represented because it is directed from Centaurus almost straight away from our line of sight where sample completion is becoming a serious issue.  However there are four rich clusters in this feature (hence the name label `4 clusters'; see the galaxies and guide line colored orange), so together with the adjacent Centaurus Cluster this part of our supercluster complex has the densest collection of clusters.  In addition to the clusters A3537, A3565, A3574, and S753 \citep{1989ApJS...70....1A} the structure extends to A3581 outside the display box of these figures at a distance of about 80 Mpc.

The last and least of the filaments extending from the Centaurus Cluster has been called the Southern Supercluster \citep{1956VA......2.1584D} and here is represented by two forks labeled SSCa (in blue) and SSCb (in a darker blue).  SSCa extends to include the Fornax Cluster and the Dorado and Eridanus clouds.  Since this structure extends to the foreground of the Centaurus Cluster it is reasonably populated in our redshift census, the inverse of the situation with `4 clusters'.  The branch SSCb is an example of an extensive but low density filament with no concentration of galaxies that warrants a cluster designation.  It is called the Telescopium-Grus Cloud in the Nearby Galaxies Atlas \citep{1987nga..book.....T}.  The most uncertain part of this structure is where it crosses the ZOA \citep{2000A&ARv..10..211K}, especially so since this is the most distant part of the structure from us and because the density of galaxies in this structure is low.  

If one gives attention to Figure \ref{5strands_yz} it is seen that the SSCb and NPI filaments lie nearly on top of each other in this projection and above left of the loci of these features the map is rather empty.  This is the region of the Local Void \citep{1987nga..book.....T}.  While it is true that the ZOA projects onto this space, it is not so wide in SGY, and the relative unimportance of ZOA losses in the bottom 60\% of the figure attest that the emptiness in the region labeled Local Void has a physical cause.  The SSCb and NPI structures are parts of the wall of this immense void \citep{2008ApJ...676..184T}.  It happens that the SSC and NPI structures are also parts of walls bounding the Sculptor Void.  It would take us too far afield to discuss this latter structure but a movie of it can be seen at a website\footnote{http://www.ifa.hawaii.edu/$\sim$tully/outreach/struc.html} by selecting `South Pole Void' (an alternative name of the Sculptor Void).

Aside from the five strands that radiate from the Centaurus Cluster there is only one other important structure in our resident supercluster complex.  A substantial filament runs from the Hydra Cluster over to the region of the Cancer Cluster \citep{1983ApJ...268...47B}.  This structure can be seen end-on in Figure \ref{V8K} and partially in Figures \ref{V8K-SGY-SGZ} and \ref{V8K-SGX-SGZ}.  It will only be mentioned here that any connection between Hydra and the other structures that have been discussed is weak.  However, with everything that has been said in this section it is to be appreciated that the structures are being followed in redshift space.  We can only arrive at a deeper understanding of the cosmography of the region by accounting for velocity flows and transforming to real space.

\section{Peculiar Velocities and Wiener Filter Reconstruction}

Distances can be measured for a fraction of the galaxies in redshift surveys.  Such measurements can be used to derive the line of sight component of peculiar velocities, $V_{pec}$
\begin{equation}
V_{pec}=V_{CMB}- d{\rm H}_{0}  
\end{equation}
where $V_{CMB}$ is a systemic velocity in the rest frame of the Cosmic Microwave Background, $d$ is distance, and H$_0$ is a value of the Hubble Constant consistent with the techniques providing distances, here taken to be H$_0 = 74$~\kmsMpc.    The distance material for the current presentation is drawn from \citet{2008ApJ...676..184T}, a compendium we call Cosmicflows-1.  This compilation of distances has the virtue that it provides dense coverage locally but the debility that it extends to only 3300 \kms.  As a consequence our current velocity-based reconstructions are restricted to the inner part of the V8k redshift cube.  This situation will be dramatically ameliorated once Cosmicflows-2 becomes available \citep{2011MNRAS.414.2005C} providing distances to 30,000 \kms.

\begin{figure}
\includegraphics[width=\textwidth]{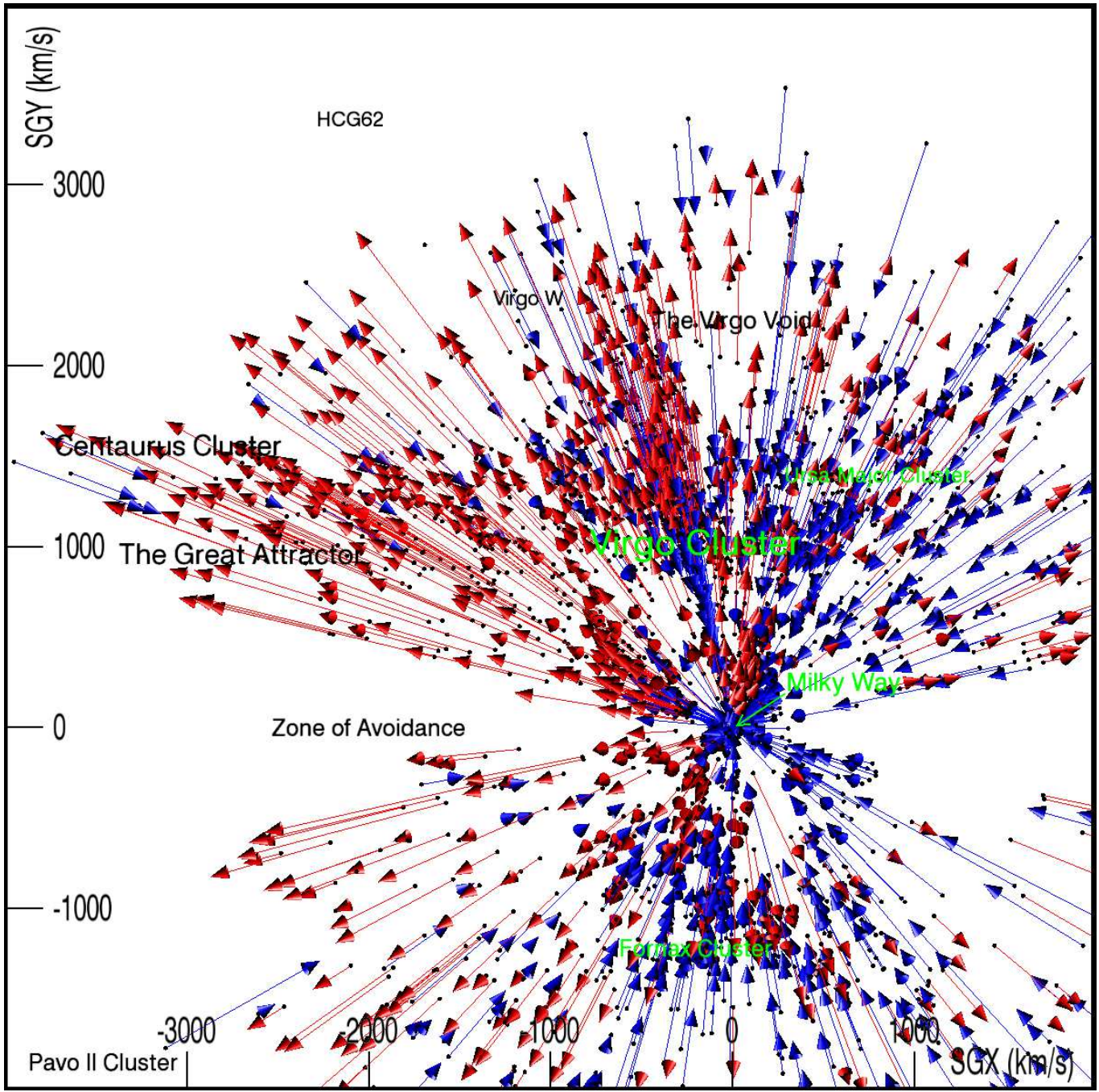}
\caption{Cosmicflows-1 peculiar radial velocities in a slice viewed in the SGX-SGY plane.  Galaxies within the interval in depth -1500 \kms\ $\leq$ SGZ $\leq$ 1500 \kms\ are located at the points. Each arrow represents a measured line-of-sight peculiar velocity with respect to the rest frame of the Cosmic Microwave Background.   Blue arrows indicate motions toward us and red arrows indicate motions away.  The amplitude of a peculiar velocity is given by the length of an arrow, with the scale matching the coordinate labels.} 
\label{CF1_XY}
\end{figure}

We return to our movie as it progresses through three sequences of information from three orthogonal views.  We begin with a pan and rotation of a slice of the V8k data cube to a supergalactic polar view (looking down from positive SGZ).  There is a dissolve to just the galaxies within the slice with measured distances and, as seen in Figure~\ref{CF1_XY}, there is a representation with blue arrows indicating motion toward us and red arrows indicating motion away from us.  The reference frame is the CMB and the length of the arrows are related to the amplitude of peculiar motion.  A great deal of coherence in peculiar velocities is evident in this figure, with many blue arrows on the right and red arrows on the left.

The next step in the analysis requires a smoothed three-dimensional reconstruction of the observed peculiar velocities that is achieved using the Wiener Filter (WF).  The WF constitutes the optimal linear  minimal variance estimator given a general data set and an assumed prior cosmological model.
The WF algorithm provides an optimal tool for the reconstruction of large scale structure from a peculiar velocity survey. The general WF  framework has been reviewed by \citet{1995ApJ...449..446Z} and \citet{2009LNP...665..565H}. The application of the WF 
to peculiar velocity observations was first presented in \citet{1999ApJ...520..413Z}. The reconstruction of the Local Universe we are displaying
in this article, based on the Cosmicflows-1 peculiar velocity catalog, is described by \citet{2012ApJ...744...43C}.

The WF constitutes a conservative estimator that is designed to find the optimal balance between the observational data and the prior model one is assuming in constructing the WF \citep{1995ApJ...449..446Z}. In regimes where the data is dense and accurate, the WF  is dominated by the data while in regimes where the data is very noisy or very sparse, like in the ZOA and towards the edge of the peculiar velocity catalog,
the WF recovers the prediction of the prior model. In the standard model of cosmology the expected mean density departures and velocity fields are zero.
In observational surveys of peculiar velocities, with increasing distance the statistical error of individual measurements increases and data becomes increasingly sparse.  The degradation in the quality of the data with the distance implies that the WF reconstruction is expected to attenuate the reconstructed fields toward the null fields.  
%This is clearly seen in Figures~\ref{WF_tomo_XY} and~\ref{WF_tomo_YZ} where the reconstructed
%velocity field represented with white arrows is null on the edges of the reconstructed box of Universe.

\begin{figure}
\includegraphics[width=\textwidth]{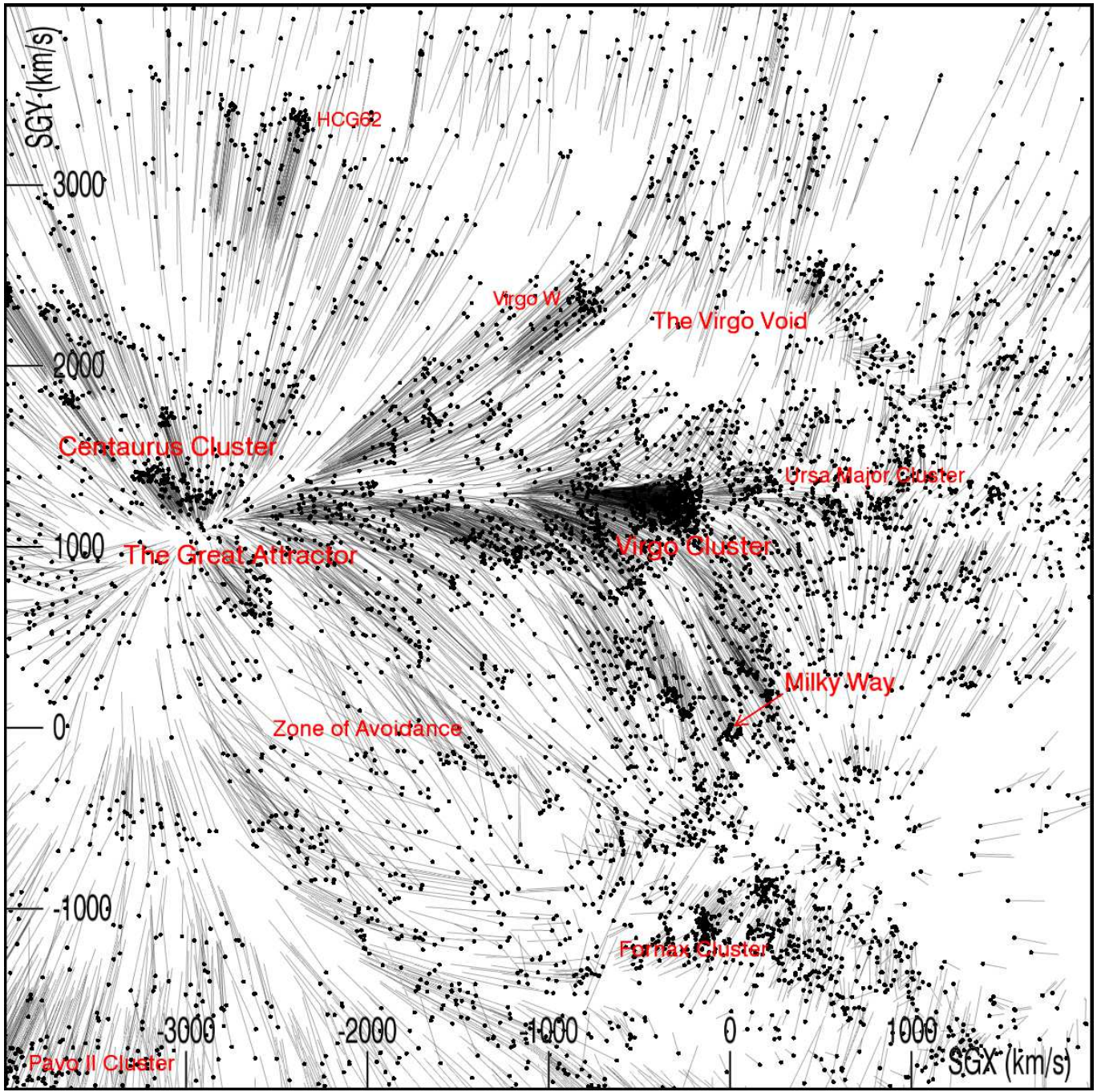}
\caption{Cosmic flows inferred from a Wiener Filter analysis of the Cosmicflows-1 sample of distances.  The view and slice are identical to that shown in the previous figure.  All V8k galaxies within the slice are shown as points with vectors indicating their motions from the Wiener Filter reconstruction.} 
\label{WF_V8K_XY_vect}
\end{figure}

In the movie there is a dissolve from observed peculiar velocities to velocities inferred from the WF analysis, the latter shown in Figure \ref{WF_V8K_XY_vect}.  A vector is attributed to every galaxy in the slice.  The input observed peculiar velocities for a fraction of the galaxies are radial components and point toward or away from the Milky Way.  The WF peculiar velocities are three-dimensional and need not align in our direction.  If one compares with the lower panels in Figure 7 of  \citet{2012ApJ...744...43C} it will be seen that we are showing here the {\it local} components of the reconstructed WF velocity field; ie, the components due to mass irregularities within the data zone.  We do not show the {\it tidal} component that must be attributed to perturbations on scales well beyond the observations.  The dominant characteristic of the flow patterns in Figure \ref{WF_V8K_XY_vect} is convergence near the location of the Centaurus Cluster and the region labelled Great Attractor.

\begin{figure}
\includegraphics[width=\textwidth]{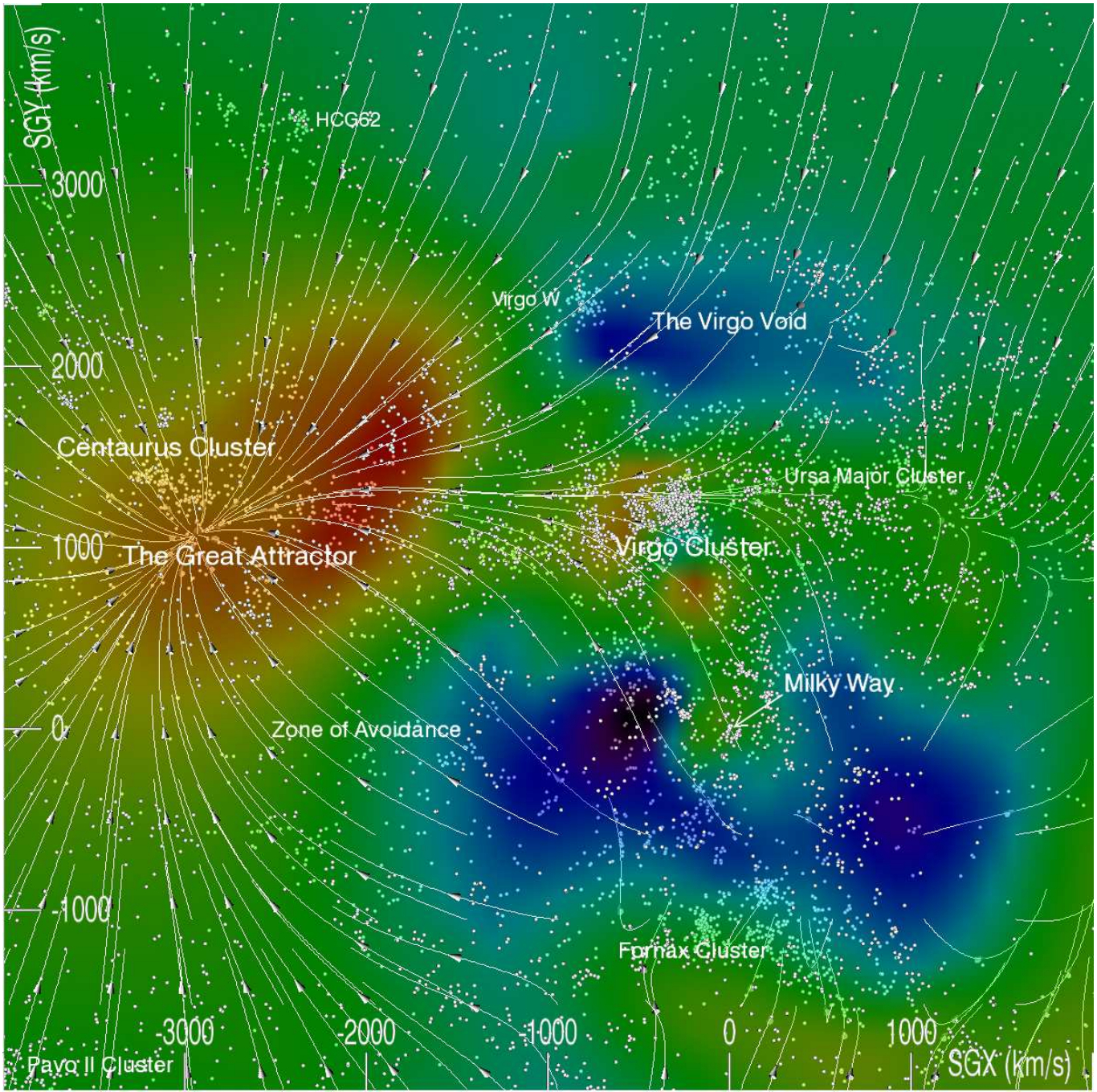}
\caption{Cosmic flows and the underlying density field from the Wiener Filter analysis.  The view and slice are identical to those shown in the previous two figures.  White dots locate galaxies from the V8k catalog that lie within the slice.  White lines identify streamlines of motion, prominent in this slice toward the Virgo Cluster and then culminating in the region of the Centaurus Cluster.  The colors provide a description of the underlying density at SGZ=0 as inferred from the Wiener Filter reconstruction, with a progression from blue in voids to red in the regions of highest density.}   
\label{WF_V8K_XY}
\end{figure}

The ensuing dissolve to Figure \ref{WF_V8K_XY} still shows the placement of galaxies and the WF reconstructed velocities, now as flow streamlines.  The visual emphasis in this figure, though, is retained in the colored contours that relate to the inferred density distribution.  The Centaurus/Great Attractor region, the focus of the flow lines, is colored red for high density while the emptiest zones are blue.  At the location of the Milky Way the flow in this projection is toward the Virgo Cluster and then bends to align with the Virgo filament (Fig.~\ref{5strands_xy}), thence on to the Centaurus Cluster.  This flow is cleanly outside the zone of avoidance.

\begin{figure}
\includegraphics[width=\textwidth]{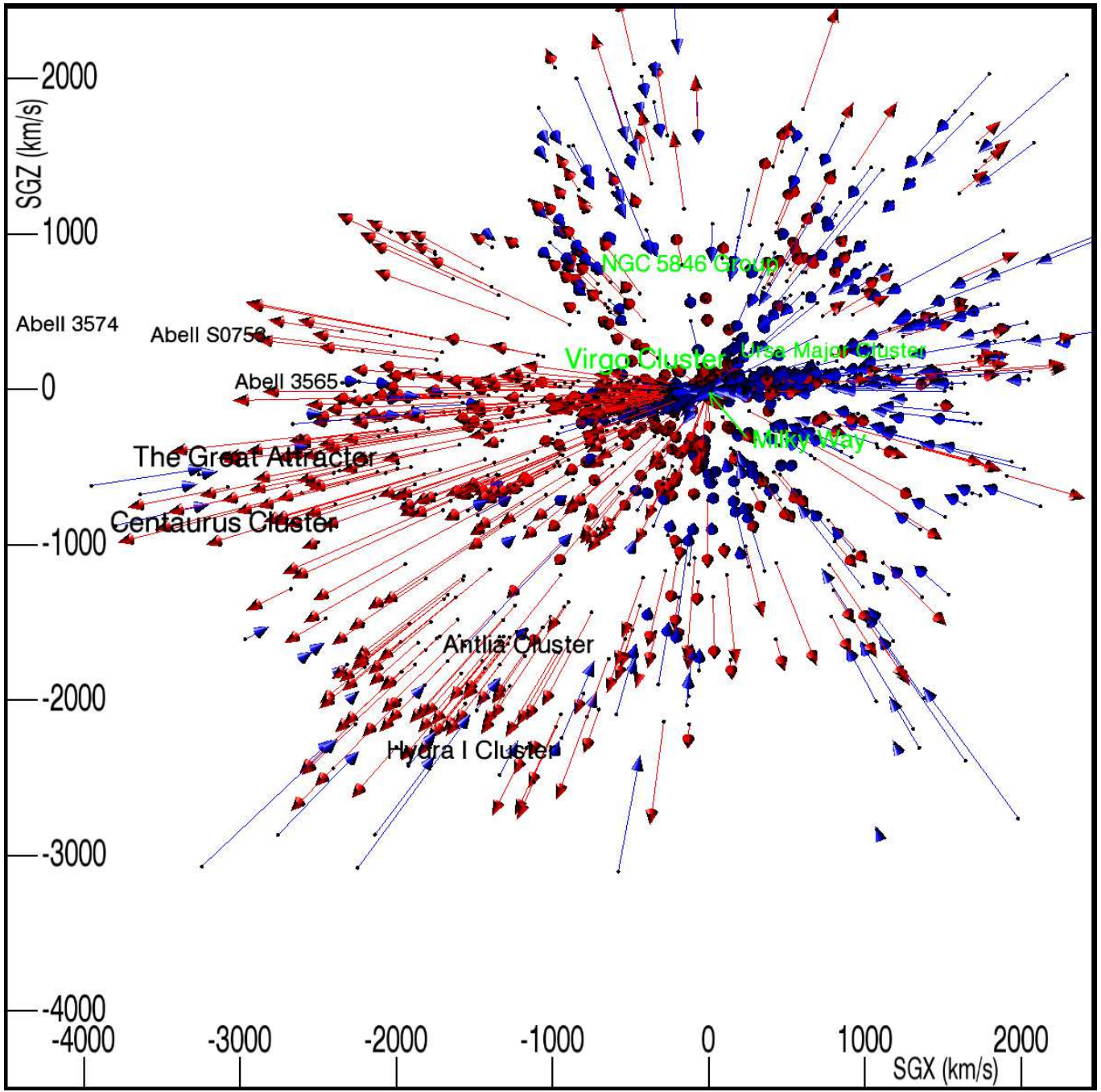}
\caption{Cosmicflows-1 peculiar radial velocities viewed in the SGX-SGZ plane centered on the Local and the Hydra-Centaurus supercluster structures.  We see galaxies with distance measures in the Cosmicflows-1 catalog that are located in a slice defined by 0 $\leq$ SGY $\leq$ 2000 \kms.  Vectors illustrate measured radial peculiar velocities in the CMB frame, blue for negative radial velocities and red for positive radial velocities.}
\label{CF1_XZ}
\end{figure}

\begin{figure}
\includegraphics[width=\textwidth]{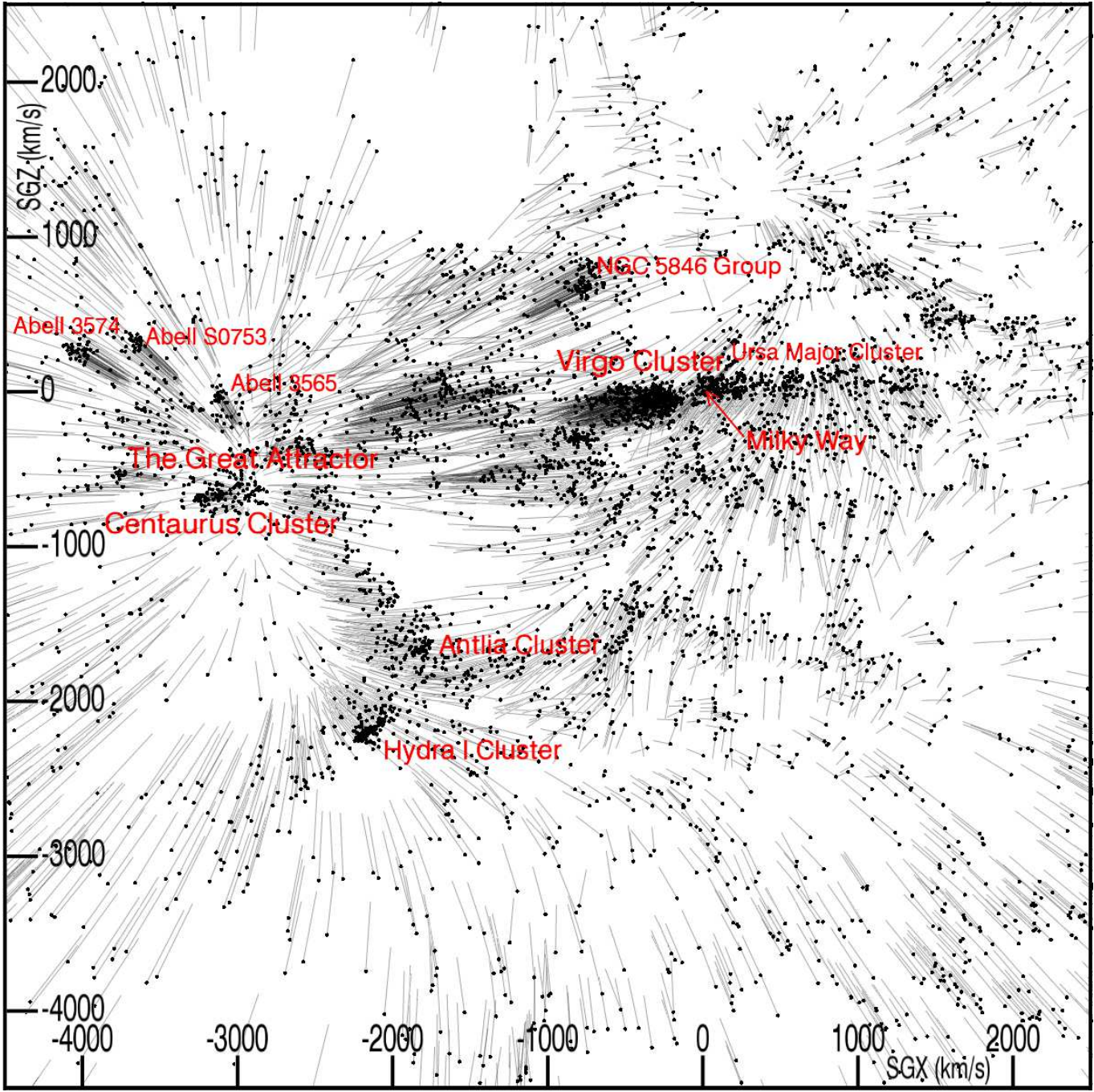}
\caption{Cosmic flows within the slice shown in the previous figure derived from the Wiener Filter reconstruction.  V8k galaxies are located by black dots and their motions from the Wiener Filter model are shown as vectors.}
\label{WF_V8K_XZ_vect}
\end{figure}

\begin{figure}
\includegraphics[width=\textwidth]{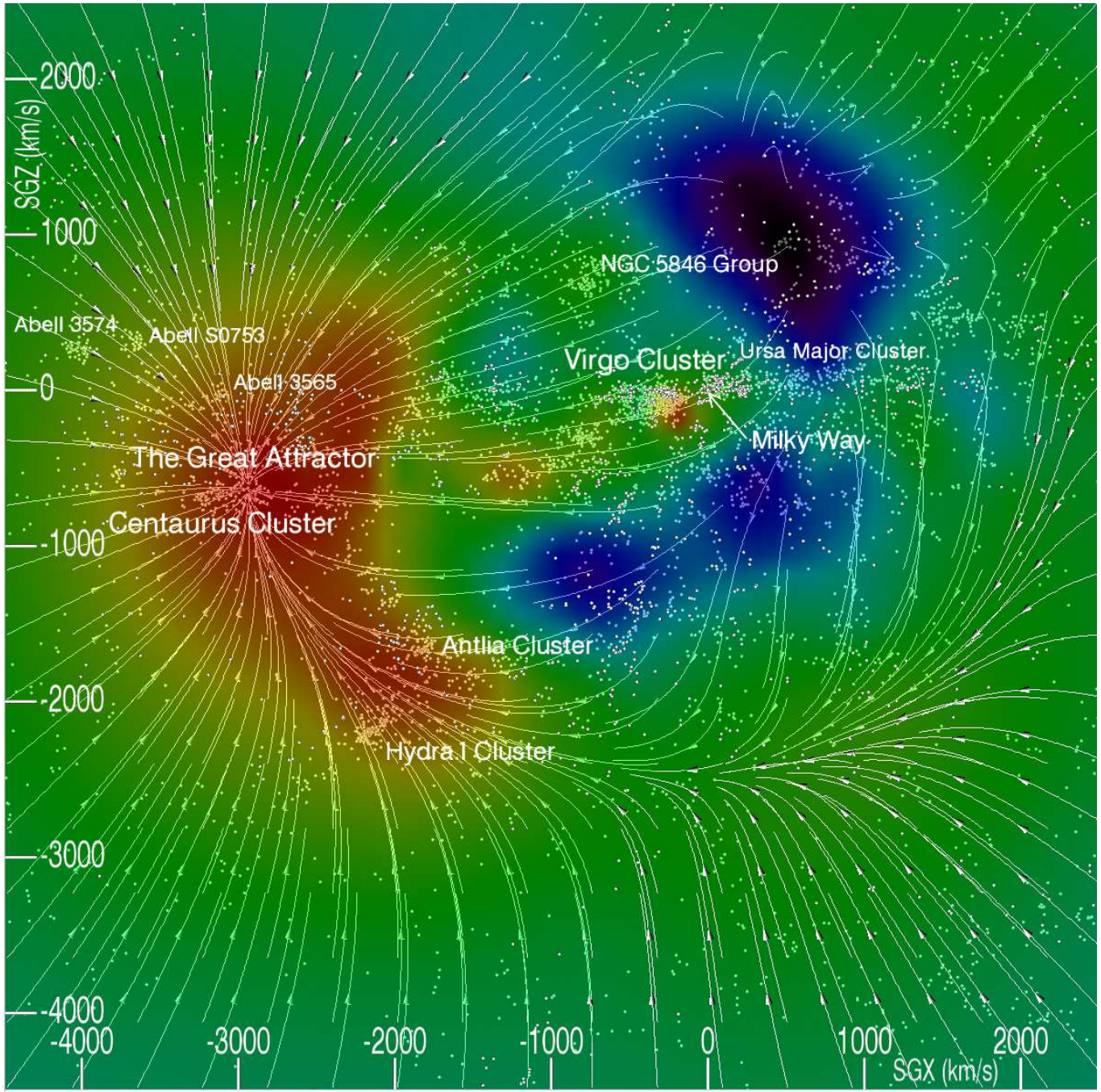}
\caption{Cosmic flows and the underlying density field from the Wiener Filter analysis for the slice shown in the previous two figures.  Colors indicate relative densities on the plane at SGY=800 \kms. The figure highlights the importance of the overdense region between the Centaurus and Hydra clusters to local dynamics.  There is a major flow through the Antlia Cluster to Centaurus and another flow along the filament from Virgo to Centaurus.}
\label{WF_V8K_XZ}
\end{figure}

These sequences are repeated with two more orientations.  With Figures \ref{CF1_XZ}, \ref{WF_V8K_XZ_vect}, and \ref{WF_V8K_XZ} the viewer looks in parallel to the negative SGY axis.  In this case the slice is restricted in the line of sight to coincide with the two nearest major filaments running into the Centaurus Cluster: those labeled `Virgo' and `Antlia' in the discussion around Figures \ref{5strands_xy}$-$\ref{5strands_yz}.   The flow up the Antlia filament toward Centaurus is particularly striking in the two WF reconstruction figures.

\begin{figure}
\includegraphics[width=\textwidth]{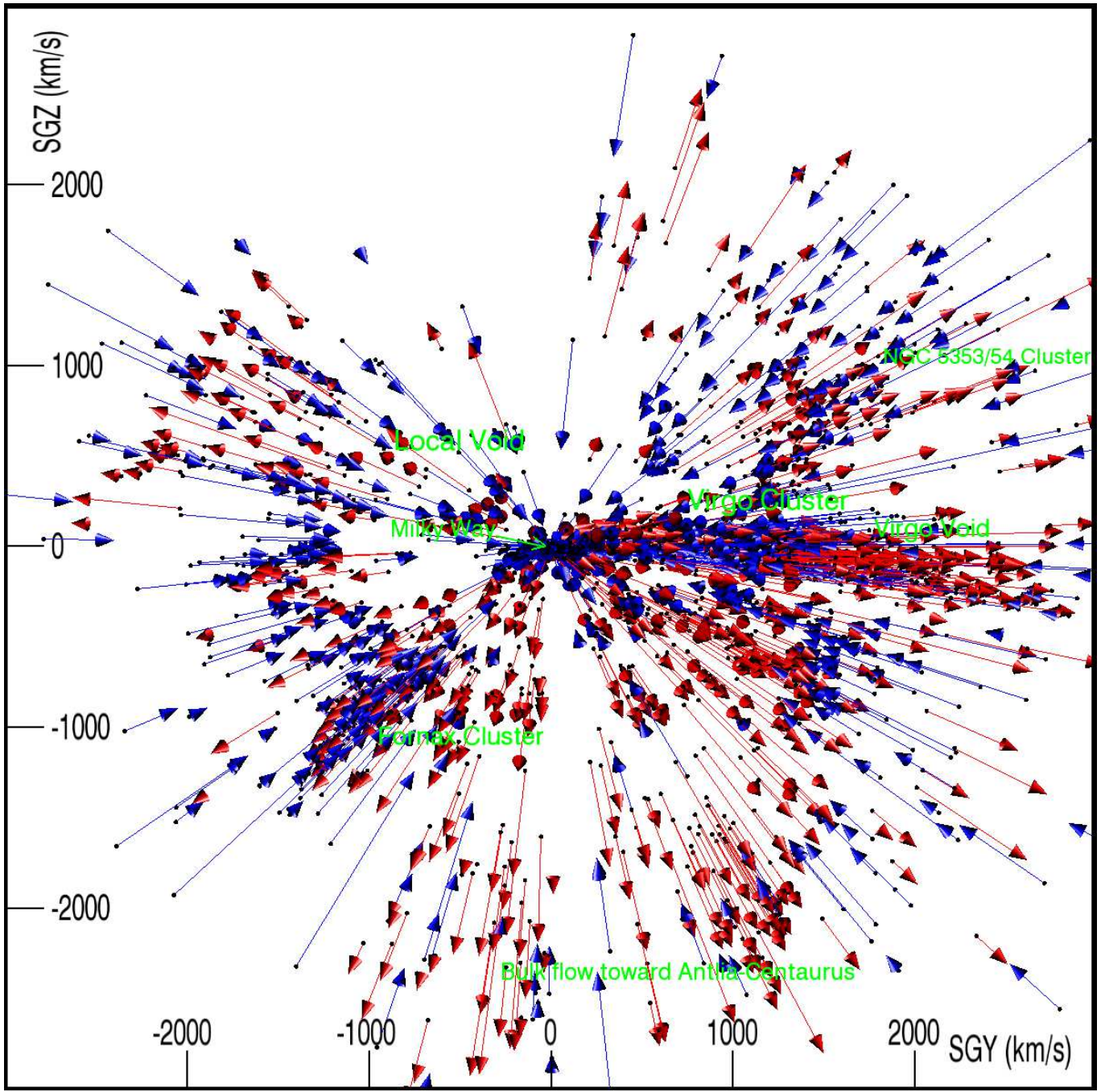}
\caption{Cosmicflows-1 peculiar radial velocities viewed in the SGY-SGZ plane, the third orthogonal viewing angle.  Galaxies with measured distances are shown if located in the slice defined by -1500 \kms\ $\leq$ SGX $\leq$ 1500 \kms.  Radial peculiar velocity vectors in the CMB frame are again shown with blue for negative velocities and in red for positive velocities.}
\label{CF1_YZ}
\end{figure}

\begin{figure}
\includegraphics[width=\textwidth]{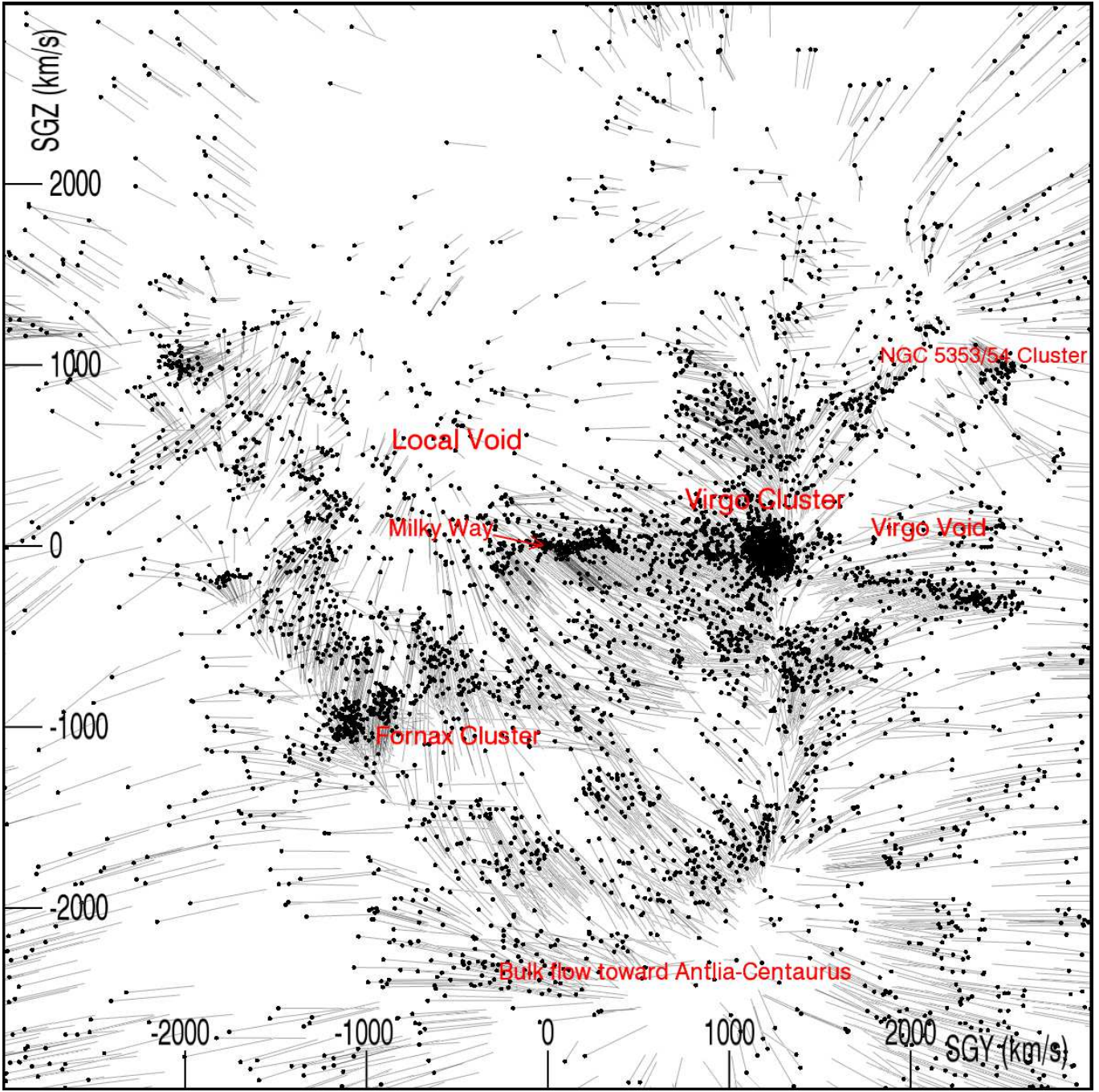}
\caption{Cosmic flows associated with the slice shown in the previous figure derived from the Wiener Filter reconstruction.  All V8k galaxies within the slice are shown as black dots and vectors show their reconstructed motions.}
\label{WF_V8K_YZ_vect}
\end{figure}

\begin{figure}
\includegraphics[width=\textwidth]{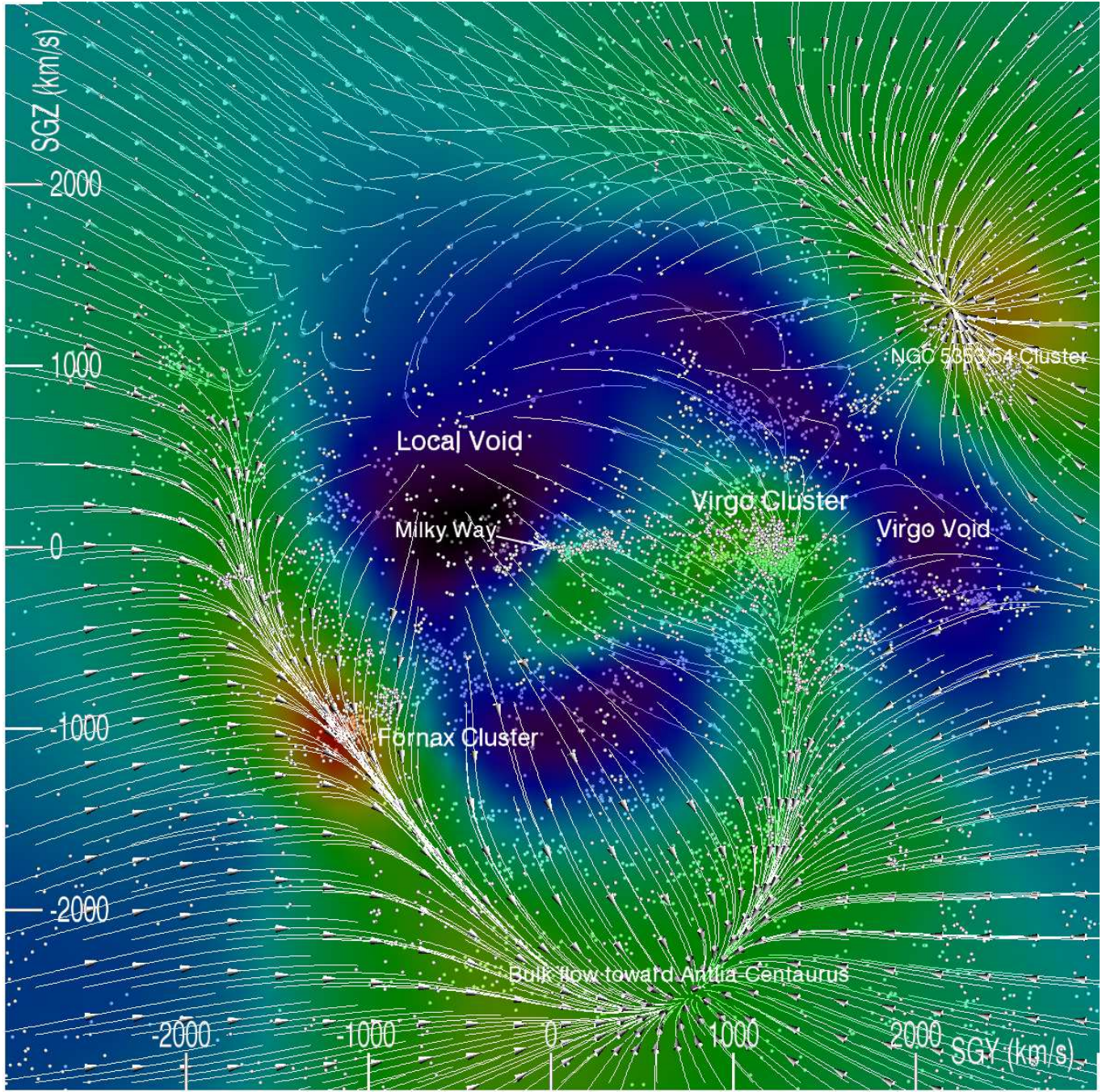}
\caption{Snoopy dog.  Cosmic flows and the underlying density field in the slice shown in the previous two figures.  This figure provides a particularly good view of the Local Void and the horse collar continuation of this void to behind the Virgo Cluster.  The flow patterns out of voids, seen in all the projections, are particularly evident in this view.}
\label{WF_V8K_YZ}
\end{figure}

With Figures \ref{CF1_YZ}, \ref{WF_V8K_YZ_vect}, and \ref{WF_V8K_YZ} ("Snoopy Dog") the viewer looks in along the positive SGX axis (as with Fig.~\ref{V8K-SGY-SGZ} but mirrored from Fig.~\ref{5strands_yz}).   This slice includes the Virgo and Fornax clusters but not the Great Attractor region to the background or the Perseus-Pisces filament to the foreground.   This slice picks up the Local Void particularly well, and shows it to be connected to the Virgo Void located directly behind the Virgo Cluster from our position at Snoopy's nose.  The WF figures in this sequence are particularly good at illustrating the streamlines of evacuation from the voids.  Not too much should be made of the apparent flow convergence points in this figure.  The slice catches flows in projection that are splitting: part at negative SGX toward Centaurus along the filaments called SSC and Antlia in Section~\ref{sec:node} and part at positive SGX of flows that we see in a preliminary version of the more extensive Cosmicflows-2 data set to be moving toward the Perseus-Pisces complex.   The high density region labeled NGC5353/54 Cluster \citep{2008AJ....135.1488T} is part of a structure called Canes Venatici- Camelopardalis Cloud in the Nearby Galaxies Atlas \citep{1987nga..book.....T}.  With the preliminary Cosmicflows-2 information It appears that the NGC5353/54 Cluster is a nodal point for flows heading, on the one hand, to the Coma Cluster, and on the other hand, to the Perseus-Pisces complex.

\begin{figure}
\includegraphics[width=\textwidth]{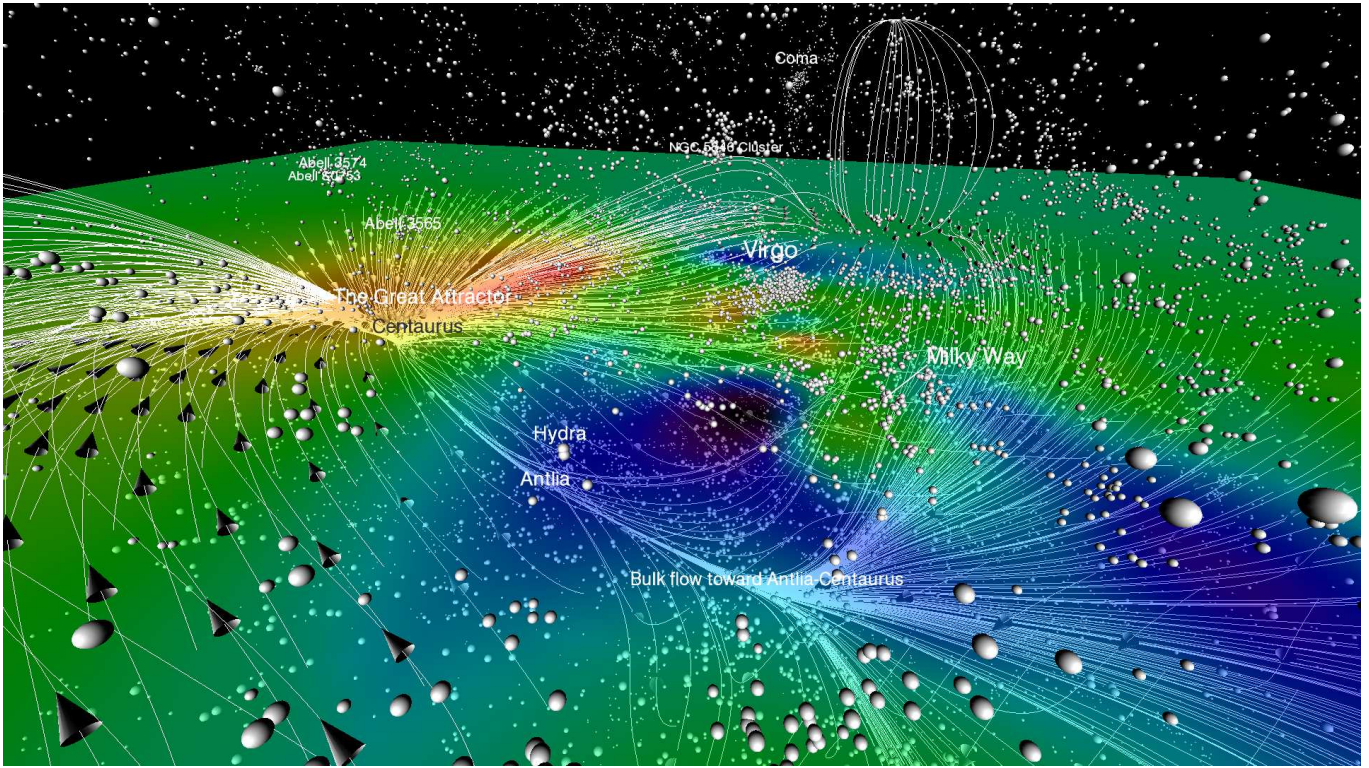}
\caption{Cosmic flows in the nearby universe. In this wide-angle perspective view, galaxies are visualized as spheres. Streamlines
of the cosmic flows are seeded on the supergalactic plane.  An image of the reconstructed overdensity source
field is also displayed on this plane. The bulk flows along the Virgo-Centaurus filament and along the Antlia-Centaurus filament are
well-defined. To the rear of the scene, beyond the Virgo Cluster, a ``fountain" is associated with the attractor also
visible in Figure~\ref{WF_V8K_YZ}.}
\label{WF_Fountain}
\end{figure}

Our movie ends with the rather fanciful Figure~\ref{WF_Fountain}.  We see the flow out of voids toward adjacent walls then the gathering of streamlines in rivers around voids that head toward the density peaks.  The `fountain' is lifting to the NGC5353/54 Cluster.  This is at the edge of the data zone so it should not be over interpreted.

\section{Summary}

We are embarked on an effort to integrate movies into discussions of the nature and development of large scale structure.  Movies have been used quite frequently to illustrate simulations.  Their use has been much more limited in connection with the real universe.  In the case of the real universe, the defining particles (galaxies) number in the tens of thousands rather than millions and they are lost from catalogs with distance, obscuration, and the vagaries of sampling over the sky.   Galaxies are only located in redshift space with any certainty and velocity perturbations in the vicinity of clusters are large.  Velocities can only be measured in the radial direction, distances have large uncertainties, and deviations from cosmic expansion are {\it very} uncertain for individual objects.

There are even more fundamental issues.  We know simulations are fake so with them we only care about statistical properties and overall features.  With observations of the real world we can be interested in specific structures because there might be grounding there for our understanding of physical processes.  We can reasonably assign names to entities to focus the discussion.  However there is the problem going in that large scale structure has three-dimensional complexity of an unknown nature.  It is an especially challenging problem to capture the wide range in structure morphologies and densities, especially where low density sheets, perhaps with spoon shapes or bifurcations, pass to greater distances or close to the ZOA.   Our own Local Sheet is an example of a low density structure that is poorly connected within the cosmic web.

It is our experience that visualization techniques are an important component of our research toolkit.  Especially, it is important to be able to navigate to different spatial perspectives or to transform between different types of information within a common scene.  To a degree, these activities can be accomplished with static images, like those included in our figures.  However, the impact is much more immediate with the movement allowed by a movie format.   A movie has at least three powerful attributes: it gives depth perspective with a three dimensional scene, it facilitates connectivity with changes in orientation or scale, and it facilitates connectivity between quite different information sets pertaining to the same physical space.

The present endeavor is preliminary.  The V8k redshift catalog is a useful proxy to the actual density field because it is locally dense, with broad representation of types, and has value-added features like structure affiliations, but in our next iteration we will give preference to the K=11.75 2MASS Redshift Survey \citep{2012ApJS..199...26H} to inform a more quantitative analysis and one that can reach to greater distances.  Of greatest importance, the next iteration will benefit from distances from the Cosmicflows-2 compendium, in preparation.   This new catalog of distances is much more extensive in number and depth than Cosmicflows-1 \citep{2011MNRAS.414.2005C}.  This improvement will have particular consequence with the Wiener Filter reconstructions. There are inevitable artifacts at the edges of the distance data zone in WF reconstructions as the calculated densities decay to the mean.  With the 3300 \kms\ limit to the Cosmicflows-1 distances the onset of these artifacts is uncomfortably close.  These data do not provide a mapping of the flows toward the adjacent structures in Perseus-Pisces and the Great Wall that we learn from the incompleteness-corrected redshift survey are surely more important than the structure that we live in.   

Nonetheless the present reconstructions are already dramatic.  The WF reconstructs the linear density field, namely the density field that 
would have existed today if the linear theory of structure formation had existed all the way to the present epoch.
Yet, there is a clear monotonic relation between the hypothetical linear density field and the actual non-linear one.
The WF reconstructions are particularly impressive in capturing the properties of the nearest voids.  There are clear signatures of outflow and expansion.  From velocity information alone, we find minor flows that converge on major filaments and then progress toward the known peaks in the distribution of galaxies around the Centaurus, Virgo, and Fornax clusters.  The agreement between the two distinct descriptions of the distribution of matter (one from redshift surveys and the other from peculiar velocity studies) is a non-trivial result.  The inference of mass distributions from observed motions presumes the hierarchical clustering theory description of the generation of those motions from the growth of gravitational perturbations.  There has been ample quantitative demonstrations of the consistency of this theory with observations \citep{1993ApJ...412....1D, 1997ApJS..109..333W, 2010ApJ...709..483L, 2011ApJ...736...93N}.
The video representations provide impressive visual support for the standard assumption of the rule of gravity. 

SDvision software has been used in this article to provide a series of visualizations of the local galaxy distribution, reconstructed matter density, and velocity fields.  The sequence of figures is an abbreviated storyboard for the accompanying movie that can be viewed at 
http://irfu.cea.fr/cosmography or http://vimeo.com/pomarede/cosmography.

\section*{Acknowledgments}   

Partiview visualization software developed by Stuart Levy at the National Center for Supercomputing Applications, University of Illinois, has been an invaluable learning tool for this project.   
We thank Martin Ratcliffe and colleagues at Sky-Skan for their support of our visualization activities.
RBT acknowledges support from the US National Science Foundation award AST09-08846. 
YH has been partially supported by the Israel Science Foundation (1013/12).
HC acknowledges support from the Lyon Institute of Origins under grant ANR-10-LABX-66.
We acknowledge the usage of the HyperLeda database (http://leda.univ-lyon1.fr).
This research has made use of the NASA/IPAC Extragalactic Database (NED) which is operated by the
Jet Propulsion Laboratory, California Institute of Technology, under contract with the National Aeronautics and Space Administration.

\bibliographystyle{mn2e}

\bibliographystyle{ApJ}
\bibliography{cosmo}

\clearpage

\clearpage

%Fig3

\clearpage
%Fig4

\clearpage

%Fig5

\clearpage

%Fig6

\clearpage

%Fig7

\clearpage

%Fig8

\clearpage

%Fig9

\clearpage
%Fig10

\clearpage
%Fig11

\clearpage
%Fig12

\clearpage

%%%%%%%%%%%%%%%%%%%%%% XY cosmicflows-1 %%%%%%%%%%%%%%%%%%%%%%%%%%%%%%%%%%%%%%%%%%%%%%%%%%%%%%%%%%%%%%%%%%%%%%%
%Fig13

\clearpage
%Fig14

\clearpage
%Fig15

\clearpage

%%%%%%%%%%%%%%%%%%%%%% XZ cosmicflows-1 %%%%%%%%%%%%%%%%%%%%%%%%%%%%%%%%%%%%%%%%%%%%%%%%%%%%%%%%%%%%%%%%%%%%%%%
%Fig16

\clearpage
%Fig17

\clearpage
%Fig18

\clearpage

%%%%%%%%%%%%%%%%%%%%%% YZ cosmicflows-1 %%%%%%%%%%%%%%%%%%%%%%%%%%%%%%%%%%%%%%%%%%%%%%%%%%%%%%%%%%%%%%%%%%%%%%%
%Fig19

\clearpage
%Fig20

\clearpage
%Fig21

\clearpage

%%%%%%%%%%%%%%%%%%%%%% 3D cosmicflows-1 %%%%%%%%%%%%%%%%%%%%%%%%%%%%%%%%%%%%%%%%%%%%%%%%%%%%%%%%%%%%%%%%%%%%%%%
%Fig22

\clearpage

\label{lastpage}

\end{document}